\documentstyle[preprint,eqsecnum,aps,epsf,floats]{revtex} 

\newcommand{\nn}{\nonumber} 

\begin{document}
\setlength\baselineskip{20pt}

\preprint{\tighten \vbox{\hbox{CALT-68-2194}
		\hbox{nucl-th/9809095} }}

\title{ Renormalization schemes and the range of two--\\[-5pt] 
nucleon effective field theory \\[5pt] }

\author{Thomas Mehen and Iain W.\ Stewart \\[10pt]}

\address{\tighten California Institute of Technology, Pasadena, CA 91125 }

\maketitle

{\tighten
\begin{abstract} 

The OS and PDS renormalization schemes for the effective field theory with
nucleons and pions are investigated.  We explain in detail how the renormalization
is implemented using local counterterms.  Fits to the NN scattering data are
performed in the $^1S_0$ and $^3S_1$ channels for different values of $\mu_R$.  
An error analysis indicates that the range of the theory with perturbative pions is
consistent with $500\,{\rm MeV}$.

\end{abstract}
}

\newpage 

\section{Introduction}

Effective field theory is an important tool for studying nuclear interactions.  To
describe low energy processes involving nucleons and pions in a model
independent way, all possible operators consistent with the symmetries of QCD
are included in an effective Lagrangian.  The short distance strong interaction
physics is parameterized by contact interactions, which can be thought of as
arising from massive states which have been integrated out.  The real power of
effective field theory is that theoretical errors can be estimated in a systematic
way.  Different contributions to an observable are organized by a power counting.
This means we have a way of organizing the theory as an expansion in
$Q/\Lambda$, where $Q$ is a momentum scale which characterizes the process
under consideration, and $\Lambda$ is the range of validity of the effective
theory.  To a given order in $Q/\Lambda$, only a finite number of operators in the
effective Lagrangian need be retained, and observables can be predicted in terms
of a finite number of experimentally determined parameters.  Theoretical
uncertainty in the calculation can be reliably estimated and reduced by
calculating higher orders in $Q/\Lambda$.  Though $\Lambda$ is not known {\it a
priori}, it is expected to be set by the masses of the lightest particles which have
not been explicitly included in the effective Lagrangian.  

In an effective field theory, ultraviolet divergences must be regulated and a
renormalization scheme defined.  The ultraviolet divergences give a constraint
on the power counting since when a divergent loop graph occurs one must
include a contact operator that can absorb the divergence at the same or lower
order in $Q$.  This is familiar from pion chiral perturbation theory. The choice of
regulator cannot affect physical results, but may make implementing a
renormalization scheme easier.  The renormalization scheme and power
counting are also tied together.  In a natural scheme, the renormalized
coefficients of the operators in the Lagrangian are normal in size  based on
dimensional analysis with $\Lambda$.  Once a power counting is established
one can translate between different renormalization schemes at a given order
in $Q$ without changing the physical predictions.  

Counting powers of $Q/\Lambda$ in the nuclear effective theory is a subtle issue
because of the large S-wave scattering length, $a$. Usually in an effective theory
(e.g., chiral perturbation theory for pions), the coefficients of operators of
dimension $n+4$ are assumed to scale as $\Lambda^{-n}$.   It is then
straightforward to examine an arbitrary graph and  determine its power in
$Q/\Lambda$. Applying this approach to the nucleon-nucleon interaction does not
work because the large scattering lengths introduce an unnatural length scale. 
Since $1/a(^1S_0)=-8.32\,{\rm MeV}$ and $1/a(^3S_1)=36.4\,{\rm MeV}$, it is
necessary to sum corrections that scale like $(Q a)^n$ to all orders. Therefore, the
power counting must be modified.  A detailed discussion of power counting in the
presence of a large scattering length is given in Refs.~\cite{W1,ksw1,ksw2,Bira}. In
Ref.~\cite{W1}, it is pointed out that the large scattering length changes the power
counting for graphs with intermediate nucleons and four-nucleon couplings with
no derivatives.  In Ref.~\cite{ksw1,ksw2}, Kaplan, Savage, and Wise (KSW), point
out that the effects of the large scattering length can be incorporated into the
theory by assigning a power counting to the coefficients of four-nucleon
operators.  The power counting for coefficients of operators mediating S-wave
transitions is changed, as well as other coefficients because of angular momentum
mixing.  A more detailed discussion of this power counting is left to the next
section.  A similar power counting is discussed in Ref.~\cite{Bira}.  

Two different calculational techniques for the effective theory of nucleons are used
in the literature.  In one approach, the power counting is applied to regulated
N-nucleon potentials and the Schroedinger equation is solved
\cite{W1,Bira2,Lepage,park}.  Solving the Schroedinger equation is equivalent to
including all ladder graphs with the potential as the two-particle irreducible kernel
(see, for e.g., \cite{ksw0}).  The second approach, advocated by KSW, is like
ordinary chiral perturbation theory in that the power counting is applied directly to
the Feynman graphs which contribute to the amplitude.  Here the method for
dealing with nuclear bound states and infrared divergences that occur at zero
kinetic energy is similar to the methods used in Non-Relativistic QED and QCD
\cite{NRQCD}.  A non-relativistic propagator is used which includes the kinetic
energy term to regulate the infrared divergence.  In the Feynman diagram approach,
dimensional regularization is the most convenient regulator, and analytic results
are readily obtained.  In the potential method, the Schroedinger equation is usually
solved numerically.  In practice, divergences are regulated and renormalized
couplings are defined using a finite cutoff scheme.  In Ref.~\cite{sf}, it has been
explicitly shown that without pions the potential method can deal with large
scattering lengths, and gives an expansion in $Q/\Lambda$.

\begin{figure}[t!]  
  \centerline{\epsfxsize=17.0truecm \epsfbox{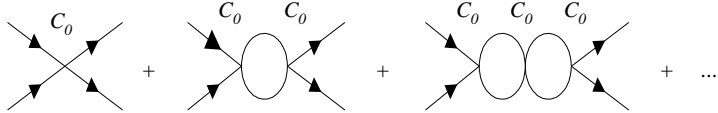}} 
{\tighten
 \caption[1]{The leading order contribution to NN scattering in the KSW power 
counting.}  \label{fig_C00} } 
\end{figure}

An important aspect of the KSW analysis is the use of a novel renormalization
scheme, power divergence subtraction (PDS).  In PDS, loop integrals in Feynman
graphs are regulated using dimensional regularization, and poles in both $d=3$
and $d=4$ are subtracted.   The subtraction of $d= 3$ poles gives a power law
dependence on the renormalization point, $\mu_R$, to the coefficients of
four-nucleon operators.   Let us use $C_{2m}$ to denote the coefficient of a
four-nucleon operator with $2m$ derivatives (where, for the moment, we restrict
ourselves to operators mediating S-wave transitions).  Choosing $\mu_R\sim Q$,
graphs with an arbitrary number of $C_0(\mu_R)$ vertices scale as $1/Q$ and must
be summed to all orders, as shown in Fig.~\ref{fig_C00}.  This is precisely the set of
graphs that sums corrections that scale like $(Qa)^n$.  Higher order contributions
form a series in $Q/\Lambda$.  In Ref.~\cite{Bira}, it is emphasized that it is possible
to phrase the power counting in a scheme independent manner.  The choice of
scheme is simply to give natural sized coefficients which make the power counting
manifest, i.e. $C_{2m}(\mu_R)\sim 4\pi/(M\Lambda^m\mu_R^{m+1})$, where
$C_{2m}(\mu_R)$ are the renormalized couplings.  PDS is one example of such a
scheme.  In Ref.~\cite{Cohen1}, it is shown how the KSW power counting can be
implemented by solving the Schroedinger equation in a finite cutoff scheme.

Pions can be added to the effective field theory by identifying them as the
pseudo-Goldstone bosons of the spontaneously broken chiral symmetry of QCD. 
All operators with the correct transformation properties are added to the effective
Lagrangian.  This includes operators with insertions of the light quark mass matrix
and derivatives, whose coefficients are needed to cancel ultraviolet divergences
from loop graphs.  In dimensional regularization, these ultraviolet divergences are
of the form $p^{2n} m_\pi^{2m}/\epsilon$.  For instance, for nucleons in the
${}^3\!S_1$ channel, the two loop graph with three pions and a two loop graph
with two pions and one $C_0$ have ultraviolet divergences of the form
$p^2/\epsilon$.  This pole must be cancelled by a counterterm involving a
four-nucleon operator with 2 derivatives.   Because divergences of the form
$p^{2n}/\epsilon$ must be cancelled by local counterterms, pion exchange can
only be calculated in a model independent way if higher derivative contact
interactions are included at the same order that these divergence
occur\cite{lm,Savage}.  In Weinberg's \cite{W1} power counting, pion exchange is
included in the leading order potential.  Therefore, graphs with arbitrary numbers of
pions are leading order, while the counterterms necessary to cancel the ultraviolet
divergences in these graphs are subleading.  However, the potential method can
still be used.  As higher order derivative operators are added to the potential the
accuracy is systematically improved, because the onset of the model dependence
of the pion summation appears at higher order in $Q/\Lambda$.  For example, the
cutoff dependence of the two pion graph with one $C_0$ will be cancelled by
cutoff dependence in $C_2$.

Different estimates of the range, $\Lambda_\pi$, of an effective theory of nucleons
with perturbative pions exist in the literature.  Some authors
\cite{Gegelia2,sf2,Cohen2} argue that $\Lambda_\pi$ is as small as $m_\pi$, so that
including perturbative pions is superfluous.    One estimate of the range is given by
KSW who conclude that $\Lambda_\pi \sim 300\,{\rm MeV}$.  They point out that in
PDS the renormalization group equation for the coefficient $C_0(\mu_R)$ is
modified by the inclusion of pions in such a way that for $\mu_R \gtrsim 300\,{\rm
MeV}$, $C_0(\mu_R)$ scales like $\mu_R^0$ instead of $\mu_R^{-1}$.  Since the
power counting is no longer manifest above this scale, KSW conclude that the
effective theory breaks down at this point.  In Ref.~\cite{sf2} different renormalized
couplings are obtained.  Here a breakdown of the power counting for
$C_2(\mu_R)$ at $\mu_R\sim m_\pi$ is observed.  A crucial question is whether a
breakdown in the running of the coupling constants is a physical effect or simply
an artifact of the renormalization scheme.  It is dangerous to draw conclusions
based on the large momentum behavior of the coupling constants because the
beta functions of the couplings are scheme dependent\footnote{\tighten This is in
contrast with dimensionless coupling constants like $g$ in QCD.  In that case the
first two coefficients of the beta function are scheme independent, so conclusions
based on the behavior of the running coupling constant at small coupling (e.g.,
asymptotic freedom) are physical.}.  In Ref.~\cite{ms0}, a momentum subtraction
scheme is introduced where the power law dependence of the coupling constants
persists even in the presence of pions, and for all values of $\mu_R > 1/a$.  This
scheme is called the OS scheme, since in a relativistic theory it might be called an
off-shell momentum subtraction scheme.  In Ref.~\cite{Gegelia1}, a similar scheme
is applied to the spin singlet channel in the theory without pions, where it is shown
to give results identical to the PDS scheme.  The OS scheme is a natural scheme
that works with arbitrary partial waves and with pions.  Thus, the range of the
validity of the effective theory is not limited by the large $\mu_R$ behavior of the
couplings.  PDS is still a useful scheme in which to calculate observables.  If one
splits $C_0(\mu_R)$ into a non-perturbative and perturbative part,
$C_0(\mu_R)=C_0^p(\mu_R) + C_0^{np}(\mu_R)$, then $C_0^{np}(\mu_R)\sim
1/\mu_R$ for all $\mu_R > 1/a$.  Once this split has been performed, it is
straightforward to establish relations between the OS and PDS schemes order by
order in perturbation theory, and any prediction for an observable will be identical
in the two schemes up to the order in $Q/\Lambda_\pi$ to which it is calculated. 
Since in both schemes there is no scale where the power counting breaks down, it
is possible that $\Lambda_\pi > 300\,{\rm MeV}$.  

Physically, one expects the effective theory to be valid up to a threshold where
new degrees of freedom can be created on-shell. For elastic nucleon scattering,
the relevant physical threshold is production of $\Delta$ resonances which occurs
at $p = \sqrt{M_N(M_{\Delta} - M_N)} = 525~{\rm MeV}$ (the S-wave channels
couple only to the $\Delta\,\Delta$ intermediate state so $p = \sqrt{2
M_N(M_{\Delta} - M_N)} = 740~{\rm MeV}$ \cite{Savage2}).  Above this scale, the
$\Delta$ must be included as an explicit degree of freedom.  Below this scale, the
$\Delta$ can be integrated out leaving an effective theory of pions and nucleons. 
Rho exchange becomes relevant at a scale, $p\sim m_\rho = 770\,{\rm MeV}$. There
is also a $N^*(1440) N$ intermediate state with a threshold of $p=685\,{\rm MeV}$. 
One might expect $\Lambda_\pi$ to be of order these thresholds.  However, there
is an intermediate scale of $300\,{\rm MeV}$ associated with short distance
contributions from potential pion exchange\footnote{\tighten The phrase ``potential
pion exchange'' will be used for a perturbative pion with energy independent
propagator.  This is sometimes called static pion exchange.}.  Using dimensional
analysis, a graph with the exchange of $n+1$ potential pions is suppressed by
$p/300\,{\rm MeV}$ relative to a graph with $n$ potential pions.  Comparison of the
size of individual graphs is scheme dependent (for example the size of graphs differ
in MS and in $\overline{\rm MS}$).  The $300\,{\rm MeV}$ scale applies only to a
subset of graphs, and may change once all graphs at a given order in $Q$ are
included in the estimate.  Therefore, $300\,{\rm MeV}$ can be taken as an order of
magnitude estimate for the range of the theory, but the actual range may be
enhanced or suppressed by an additional numerical factor.

This then motivates the important question: How does one determine the range of
the effective field theory? This is obviously a question of great practical
importance.  Theoretical arguments can only give an approximate estimate for the
range.  A good example comes from SU(3) chiral perturbation theory.  In this
strong coupling theory, it is natural to expect that the range of the theory is the
chiral symmetry breaking scale $\Lambda_\chi \sim 2\sqrt{2} f_\pi = 1200\,{\rm
MeV}$ \cite{GM}.  However, the convergence of the momentum expansion will
depend on the particular process under consideration.  For instance, in $\pi-\pi$
scattering the range of the expansion is set by the threshold for $\rho$ production,
$m_\rho = 770\,{\rm MeV}$.  In this paper, the range of the two nucleon effective
theory will be estimated using nucleon-nucleon scattering data.  Our results are
consistent with $\Lambda_\pi \sim 500\, {\rm MeV}$.  As we will explain in section
VI, the error analysis is applied to $\delta$ rather than to $p\cot\delta$ as in
Ref.~\cite{sf2}.  This range does not depend on the value of the renormalization
point chosen, and is found in both the OS and PDS schemes.  However, only
next-to-leading order calculations have been used so it is hard to estimate the
error in this value.  When higher order corrections are computed, it should be
possible to obtain a reasonably accurate estimate of the range of the two nucleon
effective field theory with perturbative pions.  This $500\, {\rm MeV}$ estimate is
based solely on the phase shift data.  Predictions for the deuteron electromagnetic
form factors are also in reasonable agreement with the data \cite{ksw3}, but do not
probe momenta greater than $200\,{\rm MeV}$, and so do not provide interesting
information regarding the range of the theory.

In section II, we review the power counting method of KSW \cite{ksw1,ksw2}, and
the PDS scheme.  The importance of being able to count factors of the large
nucleon mass in a non-relativistic effective field theory is discussed.  We review
the OS scheme,  which is compatible with the KSW power counting.  We
describe the procedure for defining the renormalized couplings using local
counterterms for each of these schemes.
  
In section III, we discuss the theory with only nucleons, where $\Lambda\sim
m_\pi$.   Local counterterms for both the PDS and OS schemes are computed. 
These counterterms are used to obtain the beta functions for the four-nucleon
operators, and we explain why the beta functions for the most relevant
operators in this theory are one-loop exact.  

The theory with nucleons and pions is analyzed in section IV.  In the ${}^3\!S_1$
channel, there are corrections to the PDS beta functions at all orders in $Q$.    As
examples, we compute the PDS beta functions for $C_0^{({}^3\!S_1)}(\mu_R)$ to
order $Q$, and for $C_2^{({}^3\!S_1)}(\mu_R)$ to order $Q^0$.  In this channel,
even in the limit $m_\pi\to 0$, there are logarithmic divergences (poles of the form
$p^2/\epsilon$ in dimensional regularization).   In the OS scheme, the ${}^3\!S_1$
beta functions can be calculated exactly.   We compute the exact beta functions
for $C_0(\mu_R)$, $C_2(\mu_R)$, and $C_4(\mu_R)$ in the OS scheme in the
${}^1\!S_0$ and ${}^3\!S_1$ channels.  In section V, the counterterms for the
coupling constant $D_2(\mu_R)$ are derived in the OS and PDS schemes.

In section VI,  we discuss why it is important to have $\mu_R$ independent
amplitudes order by order in the expansion. In the OS scheme amplitudes are
$\mu_R$ independent, while in PDS $\mu_R$ independent amplitudes can be
obtained by treating part of $C_0(\mu_R)$ perturbatively.  If this is not done then
the sensitivity to $\mu_R$ is larger than one might expect \cite{Gegelia2}, for
reasons we explain.  Fits to the data are presented for different values of $\mu_R$
and the coupling constants in both OS and PDS are shown to evolve according to
the renormalization group equations.

In section VII,  an error analysis similar to a method due to Lepage \cite{Lepage} is
used to investigate the range of the effective field theory with perturbative pions at
next-to-leading order.  Weighted fits are performed for the scattering data in both
the ${}^1\!S_0$ and ${}^3\!S_1$ channels.  Our results rule out $\Lambda_\pi \sim
m_\pi$, and are consistent  with $\Lambda_\pi \sim 500\, {\rm MeV}$.


\section{Power counting and renormalization schemes}

In this section, the KSW power counting and compatible renormalization schemes
are discussed.  The theory containing only nucleon fields is considered first. Some
notation is set, and we explain why the large nucleon mass does not affect the
power counting.  The renormalized couplings are then defined in terms of local
counterterms, and the KSW power counting for coefficients of four-nucleon
operators is reviewed.  Next, we consider the theory including pions.  We review
the power counting for potential and radiation pions, and explain the origin of the
$300\,{\rm MeV}$ scale associated with potential pion exchange. The PDS
renormalization scheme is then discussed and we introduce the OS momentum
subtraction scheme, which is also compatible with the power counting.  

Below the scale $m_\pi$, the pion can be integrated out, leaving a theory of
non-relativistic nucleons interacting via contact interactions.  The Lagrangian 
in the two nucleon sector is given by:
\begin{equation} \label{LN}
  {\cal L}_{NN} =  N^\dagger \Big[ i\partial_t + \overrightarrow\nabla^2/(2M) 
   + \ldots \Big] N - \sum_s\,\sum_{m=0}^{\infty} C^{(s)}_{2m}\,
   {\cal O}^{(s)}_{2m}   \,, 
\end{equation} 
where $M$ is the nucleon mass, and the ellipsis refers to relativistic
corrections.  ${\cal O}^{(s)}_{2m}$ is an operator with $2 m$ spatial derivatives
and four-nucleon fields, $N$. We will work in a basis in which the operators
mediate transitions between ingoing and outgoing states of definite total
angular momentum. Our notation is set up to agree with
Refs.~\cite{ksw1,ksw2}. The superscript $s$ will give the angular momentum
quantum numbers of these states in the standard spectroscopic notation,
$^{2S+1}L_J$.  If we denote the incoming and outgoing orbital angular
momentum by $L$ and $L^{\prime}$, then any operator mediating a transition
between these states must contain at least $L+L^{\prime}$ derivatives. For
states with $S=0$,  $|L-L^{\prime}| = 0$, while for states with $S=1$,
$|L-L^{\prime}|= 0~{\rm or}~2$.  For $L=L'=0$ the first few terms in the series
are
\begin{eqnarray}  \label{L2s}
 \sum_{s,m} &C^{(s)}_{2m}&\,
   {\cal O}^{(s)}_{2m}  \\  
  &=& {C_0^{(s)}} ( N^T P^{(s)}_i N)^\dagger ( N^T P^{(s)}_i N) 
   - {C_2^{(s)}\over 8} \left[ ( N^T P^{(s)}_i N)^\dagger ( N^T P^{(s)}_i 
    \:\tensor{\nabla}^{\,2} N) + h.c. \right] + \ldots  \,, \nn 
\end{eqnarray}
where the matrices $P_i^{(s)}$ project onto the correct
spin and isospin states
\begin{eqnarray}
  P_i^{({}^1\!S_0)} = \frac1{\sqrt{8}} \, (i\sigma_2) \, (i\tau_2 \tau_i ) \ ,\qquad 
  P_i^{({}^3\!S_1)} = \frac1{\sqrt{8}} \, (i\sigma_2 \sigma_i  ) \, (i\tau_2)  \ .
\end{eqnarray}
The Galilean invariant derivative in Eq.~(\ref{L2s}) is ${\tensor{\nabla}}^{\,2} =
\overleftarrow{\nabla}^2 - 2\overleftarrow\nabla\cdot \overrightarrow\nabla +
\overrightarrow{\nabla}^2$, and the ellipsis  denote contributions with more
derivatives and/or higher partial waves.  

The $C_{2 m}$ appearing in Eq.~(\ref{LN}) are bare parameters. To
renormalize the theory, the bare coupling is separated into a renormalized
coupling and counterterms as follows:
\begin{eqnarray}
  C_{2m}^{\rm bare} &=&C_{2m}^{\rm finite} - \delta^{\rm uv}C_{2m} \,,
 \qquad\quad C_{2m}^{\rm finite} =C_{2m}(\mu_R) -
   \sum_{n=0}^\infty \delta^n C_{2m}(\mu_R) \,.  \label{ctexpn}
\end{eqnarray}
Note that we divide the counterterms into two classes. The first, which have
the superscript uv, contain all genuine ultraviolet divergences. These include
$1/\epsilon$ poles, if dimensional regularization is used, or powers and
logarithms of the cutoff if a hard cutoff is used. We will also include some finite
constants (e.g., the $-\gamma + {\rm ln}(4 \pi)$ that is subtracted in
$\overline{MS}$) if this proves to be convenient for keeping expressions
compact.  By construction, these counterterms are $\mu_R$ independent, but
will depend on $C_{2m}^{\rm finite}$.  The renormalized coupling is denoted
$C_{2m}(\mu_R)$. The remaining counterterms, $\delta^nC_{2m}(\mu_R)$,
contain no ultraviolet divergences and will be referred to as the finite
counterterms.  The choice of the finite counterterms differentiates between the
schemes in our paper. An infinite number of finite counterterms are needed
because an infinite number of loop graphs are included at leading order. The 
renormalization is carried out order by order in the loop expansion. The
superscript $n$ indicates that $\delta^nC_{2m}$ is included at tree level for a
graph with $n$ loops.  When higher loop graphs are considered, the $\delta^n
C_{2m}$ counterterm takes the place of $n$ loops \cite{Collins}.  For example,
at three loops we have three loop diagrams with renormalized couplings at the
vertices, two loop diagrams with a $\delta^1 C$ counterterm, one loop 
diagrams with either one $\delta^2 C$ or two $\delta^1 C$'s, and a tree level
diagram with $\delta^3 C$.  Examples are given in section III and appendix A.

 For the nucleon theory, the kinematic part of the power counting  is very
simple \cite{W1,ksw3}.  $Q$ is identified with a typical external momentum
characterizing the process under consideration.  For instance, in elastic
nucleon-nucleon scattering $Q\sim p$, where $p$ is the center of mass
momentum\footnote{\tighten For the scattering $N(\vec q+\vec p\,) + N(\vec
q-\vec p\,) \rightarrow N(\vec q+\vec p\,') + N(\vec q-\vec p\,')$ it is useful to
define $p = \sqrt{ ME_{tot}-\vec q^2 +i\epsilon }$, where $E_{tot}$ is the total
incoming energy, and $M$ is the nucleon mass. To simplify the notation we will
work in the center of mass frame, $\vec q=0$, where $p^2 = \vec p\,^2 = \vec
p\,'\,^2 = ME$, and $E$ is the center of mass energy.  For external particles,
one can always translate between $E$ and $p$ using the equations of
motion.}.  Each nucleon propagator gives a $Q^{-2}$, each spatial derivative a
$Q$, each time derivative a $Q^2$, and each loop integration a $Q^5$.  

For non-relativistic nucleons, the scale $M$ is much larger than typical
momenta.  The reason for using a non-relativistic expansion is that each graph
will scale as a definite power of $M$.  To see this, rescale all energies, $q^0 \to
\tilde q^0 /M$, or equivalently time coordinates, $t \to M \tilde t $, so that
dimensionful quantities have the same size.  Since the measure $d^4 x\sim M$
the Lagrange density ${\cal L}\sim 1/M$.  In coordinate space our nucleon
fields scale as $N(x)\sim M^0$ (in momentum space ${\cal L}\sim M$ and
$N(p)\sim \sqrt{M}$), so from Eq.~(\ref{LN}) $C_{2m} \sim 1/M$.  With the $M$
scaling for the couplings determined we can find the scaling of any Feynman
graph.  A nucleon propagator gives one power of $M$, each momentum space
loop integration a $1/M$, and the $M$'s from external lines are cancelled by
$M$'s from the states.   For graphs that have insertions of the four-nucleon
operators $N_P=N_L+N_V-1$, where $N_P$, $N_L$, $N_V$ are the number of
propagators, loops and vertices.  Thus, at leading order in the non-relativistic
expansion any graph built out of the interactions in Eq.~(\ref{L2s}) scales as
$M^{-1}$ since $N_P-N_L-N_V = -1$.  Therefore, the $2\to 2$ scattering
amplitude ${\cal A}\sim 1/M$.  With the definition of ${\cal A}$ used here this
scaling gives a finite cross-section in the $M\to \infty$ limit which is physically
sensible.  Since all graphs scale the same way with $M$, $M$ is irrelevant to
the power counting.   Relativistic corrections are included perturbatively
\cite{ksw2,ls}, and are generally suppressed by $Q^2/M^2$ relative  to the
leading contribution to an observable.  This type of correction will not be
considered here.

In the theory with only nucleons, the only graphs relevant to $2 \to 2$ scattering
are bubble chains. Consider a graph $\cal{G}$ with $L$ loops in the
non-relativistic limit. In dimensional regularization, each loop will give a factor $M
p/4\pi$, and there are $L+1$ vertices, each giving a factor $-i C_{2 m}^{\rm finite}
p^{2 m}$. If the operator $O_{2m}$ appears $n_m$ times in the graph ($L+1 =
\sum_m n_m$) the result is:
\begin{eqnarray}
 {\cal{G}} &=& {4 \pi \over M} \prod_{m=0}^\infty \bigg( {-i M C_{2 m}^{\rm finite}
  \over  4 \pi}  \bigg)^{n_m}  p^{\,j},   \qquad\quad 
 \mbox{where} \quad  j=\sum_{m=0}^\infty 2\,m\,n_m+ L \,. \label{pscale}
\end{eqnarray}
 If one matches onto the effective range expansion in $\overline{\rm MS}$ one
finds $C_{2m}^{\rm finite} \sim 4\pi\,a^{m+1}/(M\,\Lambda^m)$ \cite{ksw0}. We 
again see that all graphs ${\cal G}$ are proportional to $1/M$, which is also true in
PDS.  The  large S-wave scattering lengths enhance the
importance of some graphs compared to the $p$ power counting.  This affects
the power counting for S-wave couplings, and through the mixing, couplings
with $L$ and/or $L' =2$ and $S=1$.  For other channels we have the usual
chiral power counting of $p$'s.  The power counting for insertions of 
four-nucleon operators is \cite{ksw3}
\begin{eqnarray} \label{pc} 
  C_{2m}^{(s)}(\mu_R)\,&{\cal O}_{2m}^{(s)}& \sim  C_{2m}^{(L-L')}(\mu_R)\,p^{2m} 
 \sim Q^{\,q(s,m)}\ ,\qquad\quad\mbox{where} \nn \\[10pt]
  q(s,m) &=& \left\{  \begin{array}{ccl}  {m-1} & & \mbox{for}\ L=L'=0 \\
  {m} & & \mbox{for}\ S=1 \ \mbox{and}\ (L,L'=0,2) \,, \mbox{ or }\ (L,L'=2,0) \\
  {m+1} & & \mbox{for}\ S=1 \ \mbox{and}\ L,L'=2,2  \\
  {2m}& &\mbox{for all other } S, L, \mbox{ and } L' \end{array}  \right.  \,.
\end{eqnarray}
With the coefficients $C_{2m}$ scaling as in Eq.~(\ref{pc}), the graph 
${\cal G}$ scales as
\begin{eqnarray}
 {\cal G}\sim Q^{\,i} \qquad\qquad \mbox{where}\quad 
   i= \sum_m  n_m\, q(s,m) \:+\:L \,.  \label{Qscale}
\end{eqnarray}
Note that the power of $Q$ is less than or equal to the power of $p$, $i \le j$. 
A useful mnemonic for this power counting is $1/a \sim Q$, however, the 
power counting is still valid for $Qa \gg 1$.

This $Q$ power counting  will be manifest in any renormalization scheme in
which the $C_{2m}(\mu_R)$ scale with $\mu_R\sim Q$ in such a way that
Eq.~(\ref{pc}) is true.  At leading order the counterterms
$\delta^nC_{2m}(\mu_R)$ will have the same $Q$ scaling as the coefficient
$C_{2m}(\mu_R)$.  These schemes may differ by contributions in
$C_{2m}(\mu_R)$ that scale with a larger power of $\mu_R/\Lambda$, since
this will not change the power counting  at low momentum.   

Let us now discuss the theory with pions. To add pions, we identify them as
the three pseudo-Goldstone bosons which arise from the breaking of chiral
symmetry, $SU(2)_L\times SU(2)_R \to SU(2)_V$.  With the pions included in
this way, we are doing an expansion in $m_\pi/\Lambda_\pi$ and
$p/\Lambda_\pi$. Note that in this theory, no matter how small $p$ is made the
expansion parameter will never be smaller than $m_\pi/\Lambda_\pi$.  This
theory still includes the four-nucleon operators in Eq.~(\ref{L2s}), but the short
distance physics parameterized by the coefficients $C_{2m}$ is different
because the pion is no longer integrated out.  In the pion theory, the short
distance $C_{2m}$ coefficients should be independent of the scale $m_\pi$. 
All the $m_\pi$ dependence  is now contained explicitly in powers of the light
quark mass matrix in the Lagrangian.

Pions will be encoded in the representation, $\Sigma=\xi^2=\exp{(2i\Pi/f)}$,
where
\begin{eqnarray}{\tighten
  \Pi = \left(  \begin{array}{cc} \pi^0/\sqrt{2} & \pi^+ \\ 
    \pi^- & -\pi^0/\sqrt{2} \end{array} \right) \,, }
\end{eqnarray}
and $f=130\, {\rm MeV}$ is the pion decay constant.  Under $SU(2)_L\times 
SU(2)_R$ the fields transform as $\Sigma \to L \Sigma R^\dagger$, $\xi \to 
L\xi U^\dagger = U\xi R^\dagger$, and $N \to U N$. The chiral covariant 
derivative is $D^\mu = \partial^\mu + \frac12(\xi\partial^\mu \xi^\dagger +
\xi^\dagger\partial^\mu \xi )$.
With pions we have the following Lagrangian with terms involving 0, 1 and 2
nucleons:
\begin{eqnarray}
 {\cal L}_\pi &=& \frac{f^2}{8} {\rm Tr}\,( \partial^\mu\Sigma\: \partial_\mu
  \Sigma^\dagger )+ \frac{f^2 w}{4}\, {\rm Tr} (m_q \Sigma+m_q
  \Sigma^\dagger) \nn \\*
 &+&
   \frac{i g_A}2\, N^\dagger \sigma_i (\xi\partial_i\xi^\dagger - 
   \xi^\dagger\partial_i\xi) N  + N^\dagger \bigg( i D_0+\frac{\vec D^2}{2M} 
   \bigg) N \label{Lpi} \\*
 &-& \sum_{s,m} C^{(s)}_{2m}\,  {\cal O}^{(s)}_{2m}   
  -{D_2^{(s)}} \omega {\rm Tr}(m^\xi ) ( N^T P^{(s)}_i N)^\dagger ( N^T 
  P^{(s)}_i N) \nn + \ldots \,.
\end{eqnarray}
Here $m^\xi=\frac12(\xi m_q \xi + \xi^\dagger m_q \xi^\dagger)$, $m_q={\rm
diag}(m_u,m_d)$ is the quark mass matrix,  $m_\pi^2 = w(m_u+m_d)$ where
$w$ is a constant,  and $g_A=1.25$ is the nucleon axial-vector coupling.   The
ellipsis in Eq.~(\ref{Lpi}) denote terms with more derivatives and more powers
of $m^\xi$.

With pions there are additional complications to the power counting
\cite{lm,ls,Savage} which are similar to those encountered in Non-Relativistic
QED and QCD \cite{NRQCD}.  The complications arise because there are
two relevant energy scales for the pions, $E_\pi \sim Q^2/M$ for potential
pions, and $E_\pi \sim Q$ for radiation pions.  When the energy integral in
loops is performed via contour integration, the graphs with potential pions
come from terms in which one keeps the residue of a nucleon propagator pole.
In these loops, the energy of the loop momentum is $\sim Q^2/M$ and the
energy dependent pieces of the pion propagator are suppressed by an
additional $Q^2/M^2$.  Nucleon propagators give a $Q^{-2}$ and the loop
integrals give $Q^5$.  There are also radiation pion graphs, in which the
residue of the pion pole is kept. In these loops, the loop energy is $\sim Q$,
the nucleon propagators give a $Q^{-1}$, and loop integrals give a $Q^4$.  We
find that adding a radiation pion to a bubble chain of contact interactions gives
an additional suppression factor $Q^{N}$, where $N$ is the number of nucleon
propagators in the radiation pion loop.  In either case, a pion propagator gives
a $Q^{-2}$, and each $\pi N N$ vertex gives a $Q$.  The combined propagator
and vertices for a single pion exchange give $Q^0$, so the pions can be
treated perturbatively \cite{ksw2}.  

In general, pion exchange gives both long and short distance contributions. 
The short distance contributions from potential pions are important since they
may limit the range of the effective field theory.  A single potential pion
exchange gives
\begin{eqnarray}
 i \frac{g_A^2}{2f^2}\: {\vec q \cdot \vec \sigma_{\alpha\beta}\ 
  \vec q \cdot \vec \sigma_{\gamma\delta} \over  \vec q\,^2 +m_\pi^2 }\ \vec 
  \tau_1\,\cdot\,\vec \tau_2  = 
 i \frac{g_A^2}{2f^2}\: \left[ {\vec q \cdot  \vec \sigma_{\alpha
  \beta}\ \vec q \cdot \vec \sigma_{\gamma\delta} \over 
  \vec q\,^2 }\ - { m_\pi^2\ \vec q \cdot \vec \sigma_{\alpha
  \beta}\ \vec q \cdot \vec \sigma_{\gamma\delta} \over 
  \vec q\,^2\ (\vec q\,^2+ m_\pi^2 )} \right] \vec \tau_1\,\cdot\,\vec \tau_2 
  \ \,, \label{ppip}
\end{eqnarray}
where the spin indices connect to nucleon fields $N^\dagger_\alpha N_\beta
N^\dagger_\gamma N_\delta$ which belong to external lines or propagators. 
The first term dominates for $\vec q^2 \gg m_\pi^2$, and can be isolated by
taking the limit $m_\pi \to 0$.  Graphs with radiation pions are suppressed by
$m_\pi/M$.   In the non-relativistic limit, with only potential pions, the only loop
diagrams are ladders.  Consider an arbitrary graph ${\cal G}$ with $n_m$ four
point vertices, $C_{2m}^{\rm finite}$, and $k$ potential pions.  For $L$ loops, this 
graph has a total of $L+1= k +\sum_m n_m$ vertices, and with $m_\pi=0$
\begin{eqnarray}
  {\cal G} &\propto& \bigg(\frac{M}{4\pi}\bigg)^{L}\:  \bigg( \frac{-i g_A^2}{2\,f^2}
   \bigg)^k\: p^{\,j}\: \prod_{m=0}^\infty (-i C_{2m}^{\rm finite})^{n_m} \ \ =\ \ 
  \frac{4\pi}{M} \:  \bigg( \frac{-i M g_A^2}{8\,\pi f^2}   \bigg)^k\: p^{\,j}\: 
  \prod_{m=0}^\infty  \bigg( \frac{-i MC_{2m}^{\rm finite}}{4\pi}\bigg)^{n_m} \nn \,, 
  \\*[5pt]
 && \qquad\qquad \mbox{where} \quad  j=\sum_{m=0}^\infty 2\,m\,n_m+L
    \label{p2scale} \,.
\end{eqnarray} 
In the ${}^1\!S_0$ channel, the relation in Eq.~(\ref{p2scale}) becomes an equality. 
The graph ${\cal G}\sim Q^{\,i}$ where $i$ is given in Eq.~(\ref{Qscale}).  The power
counting of the $\delta^{\rm uv} C_{2m}$ counterterms is determined by the need
to cancel ultraviolet divergences, and will not spoil the scaling for the renormalized
coefficients, since $i \le j$.  For graphs with only potential pions ($n_m=0$), it
appears that our expansion is in $p/(300\,{\rm MeV})$ since 
\begin{eqnarray}
   \frac{M g_A^2}{8\pi f^2} \sim (300\,{\rm MeV})^{-1}   \,.
\end{eqnarray}
Comparing the size of potential pion graphs therefore predicts a range of
$300\,{\rm MeV}$, but the size of these graphs may change depending on the
renormalization scheme (i.e., the finite subtractions).  It is not known {\em a
priori} how the contact interactions will affect the range of the effective theory. 
The scale $300\,{\rm MeV}$ is therefore an approximate estimate for the range of
the effective field theory with perturbative pions.  A further discussion of this issue
will be taken up in section IV.

Next, consider the power counting for coefficients that multiply operators with
powers of $m_q$.  If we are interested in momenta of order $m_\pi$, then one
counts $m_q \sim m_\pi^2 \sim Q^2$.  Therefore, any interaction term that has
an operator with a total of $2m$ powers of $p$ and $m_\pi$ will scale as
$Q^{\,q(s,m)}$ where $q(s,m)$ is given in Eq.~(\ref{pc}).  For example,
$D_2^{({}^1\!S_0)}m_\pi^2 \sim Q^0$.  It is important to understand that in the
KSW power counting $D_2$ should be treated perturbatively even though the
structure of the operator it multiplies is similar to that of the leading 4 nucleon
operator with no derivatives.  Graphs with radiation pions will also give
contributions with powers of $m_\pi^2$.  In Ref.~\cite{W1}, the power counting
for a general radiation pion graph is worked out; the only change is the power
counting for the coefficients in Eq.~(\ref{pc}).  
  
\subsection{PDS}

PDS is one scheme in which the KSW power counting is manifest.  In PDS, we
first let $d=4$ and take the $\delta^{\rm uv}C_{2m}$ counterterms to subtract
$1/\epsilon$ poles as in $\overline{\rm MS}$.  We use the notation $\mu_R$ for the
renormalization point, and $\mu$ for the dimensional regularization parameter. 
In PDS, like in the $\overline{\rm MS}$ scheme, one takes $\mu=\mu_R$.  In a
momentum subtraction scheme this is not necessary.  The next step in PDS
is to take $d=3$ and define the finite counterterms, $\delta^nC_{2m}(\mu_R)$,
to subtract the $1/(d-3)$ poles in the amplitude.   Graphs which contribute are
those whose {\em{vertices}} have a total of $2m$ derivatives.  When calculating
the $\delta^nC_{2m}(\mu_R)$ we take $m_\pi=0$ since, for instance,
counterterms proportional to $m_\pi^2$ renormalize coefficients like
$D_2(\mu_R)$.   After making these subtractions everything is continued back
to four dimensions.  It is this second set of finite subtractions that gives the
right power law dependence on $\mu_R$.  To define the coefficients that
multiply operators with powers of $m_q$, a similar procedure is followed
except we count the powers of $m_\pi^2$ at the vertices.  In PDS with just
nucleons, all the graphs that affect the running of $C_{2m}(\mu_R)$ are order
$Q^{\,q(s,m)}$, except for those with intermediate states of different orbital
angular momentum. For example, the beta function for $C_4^{({}^3\!S_1)}$ has
contributions $\sim Q$ ($q({}^3\!S_1,4) = 1$), as well as contributions $\sim
Q^3$ from graphs with two $C_2^{({}^3\!S_1- {}^3\!D_1)}$ vertices.  When
pions are included there are additional graphs that are sub-leading in the
power counting and effect the running of the couplings.  In fact, in section IV
we will show that there will be corrections to the PDS beta function for
$C_0^{({}^3\!S_1)}(\mu_R)$ at all orders in $Q$.  

\subsection{OS}

Another renormalization scheme that can be used to reproduce the power
counting in Eq.~(\ref{pc}) is a momentum subtraction scheme.  A simple
physical definition for the renormalized couplings can be made by relating the
couplings to the amplitude evaluated at the unphysical momentum
$p=i\mu_R$.  This scheme will be called the OS scheme, since in a
relativistic field theory this would be referred to as an off-shell momentum
subtraction scheme.  We start by dividing up the full amplitude as
\begin{eqnarray}
  i\,{\cal A}^s =  i\, \sum_{m=0}^\infty  {\cal A}^s_{2m} +\ldots \,. \label{Asplit}
\end{eqnarray}
Here ${\cal A}^s_{2m}$ contains the Feynman diagrams that will be used to
define the coupling $C_{2m}^{(s)}(\mu_R)$ (or equivalently the counterterms
$\delta^nC_{2m}$).  The ellipsis in Eq.~(\ref{Asplit}) denotes pieces that vanish
as $m_\pi\to 0$ which are not needed to define $C_{2m}(\mu_R)$. $A_{2m}^s$ 
is defined to contain the remaining graphs that scale as $Q^{\,q(s,m)}$, where
$q(s,m)$ is defined in Eq.~(\ref{pc}).  The definition for the renormalized coupling 
is then 
\begin{eqnarray}
 i  \left.{\cal A}^s_{2m} \right|_{\mbox{\scriptsize$\begin{array}{c} \ p=i\mu_R
 \\[-4pt] \!\!\!\! m_\pi=0 \end{array}$}} = -i  C_{2m}^{(s)}(\mu_R) \:
  (i\mu_R)^{2m}\,.  \label{rc2m}
\end{eqnarray} 
As we will see, this ensures that $C_{2m}(\mu_R)$ scales in the desired way.  In
general, there may be divergent graphs scaling as $Q^{\,i}$ and $p^{2m}$ ($i \le
2m$) whose $1/\epsilon$ poles need to be absorbed by a $\delta^{\rm uv}C_{2m}$
counterterm.  For example, consider the graph with two pions and one $C_0$
shown in row four of Fig.~\ref{fig_2ppi}.  This graph has a $p^2/\epsilon$ pole
which is cancelled by a counterterm $\delta^{\rm uv}C_2$. The finite part of this
graph is used in Eq.~(\ref{rc2m}) to define $C_4(\mu_R)$ because the graph is
order $Q$.  The key point is that since $q(s,m)\le 2m$, an ultraviolet divergence
that appears in a graph of a given order can always be absorbed into a coefficient
that appeared at the same or lower order in the power counting.  Therefore, we will
define $\delta^{\rm uv}C_{2m}$ in $\overline{\rm MS}$ to subtract all four
dimensional $1/\epsilon$ poles so that these subtractions are independent of the
renormalization point.  The finite counterterms are then fixed by the renormalization
condition in Eq.~(\ref{rc2m}).

\begin{figure}[t!]  
  \centerline{\epsfxsize=17.0truecm \epsfbox{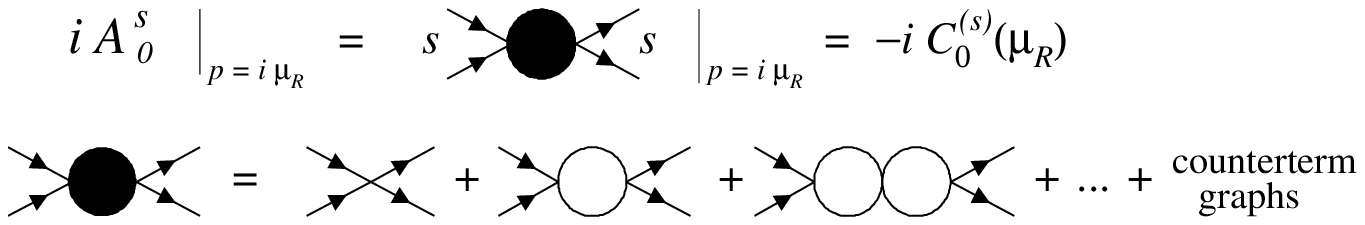}} 
{\tighten
 \caption[1]{Renormalization condition for $C_0(\mu_R)$.  The amplitude
 $i\,{\cal A}^{s,0}$ is the full four point function with $C_0^{(s)}(\mu_R)$
 and $\delta^n C_0^{(s)}(\mu_R)$ vertices, evaluated between incoming and 
 outgoing states $s={}^1S_0\mbox{ or }{}^3S_1$. }  \label{fig_C0} } 
\end{figure}

In the OS scheme, the coupling $C_0(\mu_R)$ is defined by the renormalization
condition in Fig.~\ref{fig_C0} (with or without pions).
This condition is to be imposed order by order in the loop expansion so that
graphs with $n$ loops determine $\delta^nC_0(\mu_R)$.  The $m_\pi=0$
part of pion graphs will also contribute to $C_{2m}(\mu_R)$ for $m\ge1$ in
which case the condition $m_\pi=0$ in Eq.~(\ref{rc2m}) is important.  In the
theory with pions, we also need to define couplings multiplying powers of
$m_q$, like $D_2$ in Eq.~(\ref{Lpi}).  To define these couplings we will not
include all the terms in the amplitude proportional to $m_\pi^2$.  In particular,
pion exchange graphs give long distance non-analytic contributions which will
not be used to define the running of the short distance coupling
$D_2(\mu_R)$.  The idea that long distance physics must be excluded 
from the short distance coefficients is discussed in Ref.~\cite{sf2}.
A detailed discussion of how we define $D_2(\mu_R)$ in the OS scheme will 
be left to section V.

Note that in the OS scheme there is another approach for calculating an
amplitude in terms of renormalized couplings. One can calculate all loop
graphs in ${\cal A}^s_{2m}$ in terms of the finite (or $\overline {\rm MS}$)
parameters and then demand that the renormalization condition in
Eq.~(\ref{rc2m}) is satisfied. This gives expressions for the renormalized
couplings in terms of the constants $C_{2m}^{\rm finite}$.  The amplitude can
then be written in terms of renormalized couplings by inverting these
equations.  This simplifies higher order calculations.

In the OS scheme, when an amplitude is written in terms of renormalized couplings
it will be explicitly $\mu_R$ independent at each order in $Q$.  The $\mu_R$
dependence in PDS with pions is cancelled by higher order terms.  It is possible to
obtain $\mu_R$ independent amplitudes in PDS if part of $C_0(\mu_R)$ is treated
perturbatively\cite{ms0}.  Consequences of this $\mu_R$ dependence will be
discussed in section VI.  In section III, we will see that for the theory with just
nucleons the OS scheme gives very similar definitions for the renormalized
couplings to those in PDS.  In section IV, we investigate the running couplings in
both schemes in the theory with pions.


\section{Theory with pions integrated out}

In this section, we compute the renormalized couplings in the non-relativistic
nucleon effective theory without pions.  We expect $\Lambda\sim m_\pi$.  This
theory will be examined in both PDS and the OS scheme.  The renormalization
program is implemented by explicitly calculating the local counterterms.  In
Ref.~\cite{Gegelia1}, it is shown that the PDS and OS schemes give the same
renormalized coupling constants in the $^1S_0$ channel.  Here we also consider
the spin-triplet channel and higher derivative operators.  Divergences in loop
integrals are regulated using dimensional regularization.  For the OS scheme, the
same renormalization program can be carried out using a momentum cutoff
regulator as shown in appendix A.  Following Ref.~\cite{ksw2}, we will multiply
each loop integral by $(\mu/2)^{(4-d)}$, and define $d=4-2\epsilon$.  Since there
are no logarithmic divergences in the nucleon theory, $\delta^{\rm uv}C_{2m}=0$ in
dimensional regularization.

\begin{figure}[t!]  
 \epsfysize=7.0truecm \epsfbox{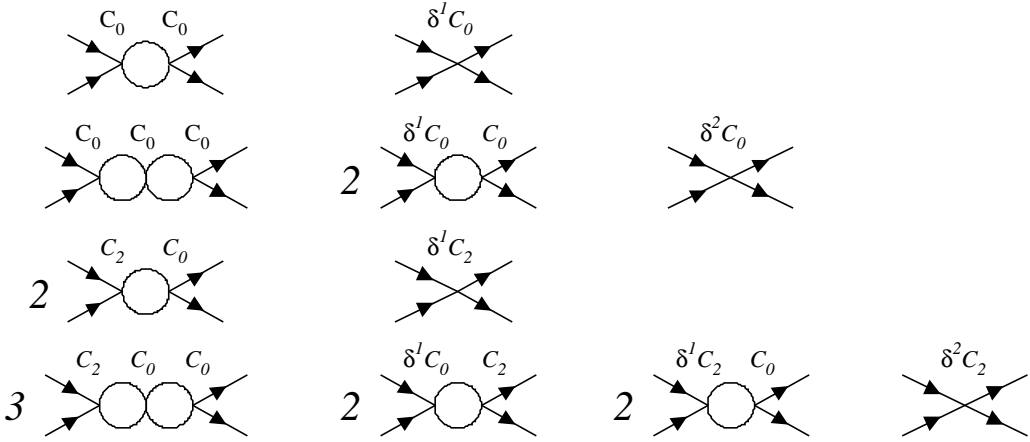}  
{\tighten  \caption[1]{One and two loop counterterms for $C_0$ and $C_2$. 
The solid lines are nucleon propagators, and symmetry factors are shown 
explicitly. The generalization to higher loops is straightforward.}   \label{fig_ct0} }
\end{figure}

In both PDS and OS scheme, it is straightforward to derive the finite
counterterms, $\delta^n C_{2m}(\mu_R)$.  The tree level graphs with
$C_0(\mu_R)$ and $C_2(\mu_R)$ satisfy the renormalization condition in
Eq.~(\ref{rc2m}).  Therefore, in both PDS and OS, $\delta^0 C_0 = \delta^0 C_2
= 0$.  At one and two loops we have the graphs in Fig.~\ref{fig_ct0}.
In $d$ dimensions, the two graphs in the first row give
\begin{eqnarray}  \label{2c0}
 && (-i C_0)^2 \Big(\mbox{\small$\frac{-iM}{4\pi}$}\Big)
  \Gamma(\mbox{\small$\frac{3-d}2$}) \Big(\frac{\mu}2\Big)^{4-d}
  \bigg( \frac{-p^2-i\varepsilon}{4\pi} \bigg)^{\frac{d-3}2}  + i\:\delta^1 C_0 \,,
\end{eqnarray}
determining $\delta^1C_0$.  In PDS, we define the counterterm to cancel the $d=3$
pole in Eq.~(\ref{2c0}) and then continue back to four dimensions.  In the OS
scheme, we take $d=4$ and demand that the contribution to the amplitude in
Eq.~(\ref{2c0}) satisfies the condition in Fig.~\ref{fig_C0}.  The counterterms
calculated in each scheme are the same (with $\mu=\mu_R$ in PDS).  In both
schemes the counterterms determined from the graphs in Fig.~\ref{fig_ct0} are
\begin{eqnarray}  
 \delta^1 C_0(\mu_R) &=& \bigg(\frac{M\mu_R}{4\pi}\bigg)\,C_0(\mu_R)^2 \ , 
   \qquad\qquad\quad
 \delta^2 C_0(\mu_R) =-\bigg(\frac{M\mu_R}{4\pi}\bigg)^2\,C_0(\mu_R)^3 \ ,\\
 \delta^1 C_2(\mu_R) &=& 2 \bigg(\frac{M\mu_R}{4\pi}\bigg)\,C_2(\mu_R)
   C_0(\mu_R)\ ,\qquad
 \delta^2 C_2(\mu_R) = -3 \bigg(\frac{M\mu_R}{4\pi}\bigg)^2\, C_2(\mu_R)
   C_0(\mu_R)^2\ . \nn
\end{eqnarray}
Note that it is essential that loop graphs also have vertices with insertions
of the counterterms.  For instance, the contribution to the amplitude from all
the graphs in the second row of Fig.~\ref{fig_ct0} is
\begin{eqnarray}
   -i C_0(\mu_R)^3  \bigg( \frac{M(ip+\mu_R)}{4\pi} \bigg)^2 \,.
\end{eqnarray}
If the one-loop graph with a $\delta^1 C_0$ counterterm had been left out then
the answer would have been proportional to $(p^2 + \mu_R^2)$ which is not
correct.  Since the loops in the nucleon theory factorize, the renormalized
$n$-loop graph gives $(ip+\mu_R)^n$.  Loop graphs will not always factorize
once pions are included.  

It is straightforward to extend this calculation to $n$ loops and to include 
higher derivatives.  In both the OS and PDS schemes, this gives the following 
counterterms ($s= {}^1\!S_0, {}^3\!S_1$, $n\ge 1$):
\begin{eqnarray}\label{ct024}
\lefteqn{{}^1S_0:\ \ \ \  } \nn \\
 &&\qquad\qquad \delta^nC_0^{({}^1\!S_0)}(\mu_R) = (-1)^{n+1}\,\left(
 {M\mu_R\over 4\pi}  \right)^n \  C_0^{({}^1\!S_0)}(\mu_R)^{\,n+1} \,, \nn \\
 &&\qquad\qquad  \delta^nC_2^{({}^1\!S_0)}(\mu_R) = (-1)^{n+1}\,\left( 
  {M\mu_R\over 4\pi} \right)^n\:  (n+1)\  C_0^{({}^1\!S_0)}(\mu_R)^{\,n} \: 
  C_2^{({}^1\!S_0)}(\mu_R) \,, \nn \\ 
 &&\qquad\qquad  \delta^nC_4^{({}^1\!S_0)}(\mu_R) = (-1)^{n+1}\,\left( 
   {M\mu_R\over 4\pi} \right)^n (n+1) C_0^{({}^1\!S_0)}(\mu_R)^{n-1} \nn \\
 &&\qquad\qquad  \qquad\qquad \times \Big[ C_4^{({}^1\!S_0)}(\mu_R)\: 
  C_0^{({}^1\!S_0)}(\mu_R) + \frac{n}2 C_2^{({}^1\!S_0)}(\mu_R)^2  \Big] \,, \nn \\  
\lefteqn{{}^3\!S_1,{}^3\!D_1:\ \ \ \  }  \\
 &&\qquad\qquad  \delta^nC_0^{({}^3\!S_1)}(\mu_R) = (-1)^{n+1}\,\left( 
  {M\mu_R\over 4\pi} \right)^n \  C_0^{({}^3\!S_1)}(\mu_R)^{\,n+1} \,, \nn \\
 &&\qquad\qquad  \delta^nC_2^{({}^3\!S_1)}(\mu_R) = (-1)^{n+1}\,\left( 
  {M\mu_R\over 4\pi} \right)^n\:  (n+1)\  C_0^{({}^3\!S_1)}(\mu_R)^{\,n} \: 
  C_2^{({}^3\!S_1)}(\mu_R) \,, \nn \\ 
 &&\qquad\qquad  \delta^nC_2^{({}^3\!S_1- {}^3\!D_1)}(\mu_R) = (-1)^{n+1}\,
   \left( {M\mu_R\over 4\pi} \right)^n \: C_0^{(^3S_1)}(\mu_R)^{\,n} \:
   C_2^{({}^3\!S_1- {}^3\!D_1)}(\mu_R)  \,, \nn \\ 
 &&\qquad\qquad  \delta^nC_4^{({}^3\!D_1)}(\mu_R) = (-1)^{n+1}\,\left( 
  {M\mu_R\over 4\pi} \right)^n C_0^{(^3S_1)}(\mu_R)^{\,n-1} \, \Big[ 
  C_2^{({}^3\!S_1- {}^3\!D_1)} (\mu_R)\Big]^2 \,. \nn 
\end{eqnarray}
Note that with $\mu_R \sim Q$, the counterterms have the same $Q$ scaling as
their corresponding coupling constant.  In the PDS scheme, there are also 
subleading terms that come from the mixing of angular momentum 
states.  In PDS
\begin{eqnarray}
 \delta^nC_4^{({}^3\!S_1)}(\mu_R) &=& (-1)^{n+1}\,\left( {M\mu_R\over 4\pi} 
   \right)^n C_0^{({}^3\!S_1)}(\mu_R)^{n-1}\Big[ (n+1)C_4^{({}^3\!S_1)}(\mu_R)\: 
  C_0^{({}^3\!S_1)}(\mu_R) \nn\\
   &&\qquad\qquad\qquad\qquad+ \frac{n(n+1)}2 C_2^{({}^3\!S_1)}(\mu_R)^2 
  +  n C_2^{({}^3\!S_1- {}^3\!D_1)}(\mu_R)^2 \Big]    \,, 
\end{eqnarray}
where the last term is suppressed by $Q^2$.  In the OS scheme
\begin{eqnarray}
 \delta^nC_4^{({}^3\!S_1)}(\mu_R) &=& (-1)^{n+1}\,\left( {M\mu_R\over 4\pi} 
   \right)^n  C_0^{({}^3\!S_1)}(\mu_R)^{n-1} \nn\\
  &&\qquad \times   \Big[ (n+1) C_4^{({}^3\!S_1)}(\mu_R)\: 
  C_0^{({}^3\!S_1)}(\mu_R) + \frac{n(n+1)}2 C_2^{({}^3\!S_1)}(\mu_R)^2  \Big] \,,  
\end{eqnarray}
which is the same as the ${}^1\!S_0$ channel.  In the OS scheme, graphs with
two $C_2^{({}^3\!S_1- {}^3\!D_1)}$ couplings and any number of
$C_0^{({}^3\!S_1)}$'s contribute to the beta function for $C_8^{({}^3\!S_1)}$
since they are order $Q^3$.  One might also ask about channels where the
large scattering length does not effect the power counting.  In this case
$C_{2m}^{(s)}(\mu_R) \sim Q^0$, and we recover the usual chiral power 
counting.   In our OS scheme, the counterterms $\delta^nC_{2m}^{(s)}(\mu_R)$ 
in these channels are either zero or a constant independent of $\mu_R$.

From Eq.~(\ref{ctexpn}) one can derive the beta functions using
\begin{eqnarray}
  \beta_{2m} \equiv \mu_R { \partial \over \partial {\mu_R} } C_{2m}(\mu_R)
  = \sum_{n=0}^\infty \mu_R { \partial \over \partial {\mu_R} } \delta^n C_{2m}
  (\mu_R) \,.  \label{beta2m}
\end{eqnarray}
The first few beta functions are 
\begin{eqnarray} \label{bet02}
\lefteqn{{}^1\!S_0:\ \ \ \  } \nn \\
  &&\qquad\quad \beta_0^{({}^1\!S_0)} = \bigg( {M\mu_R\over 4\pi} \bigg) 
  C_0^{({}^1\!S_0)}(\mu_R)^2 \,, 
 \qquad\quad \beta_2^{({}^1\!S_0)} =2 \bigg( {M\mu_R\over 4\pi} \bigg) 
  C_0^{({}^1\!S_0)}(\mu_R)\: C_2^{({}^1\!S_0)}(\mu_R)\,, \nn \\
  && \qquad\quad \beta_4^{({}^1\!S_0)} =  \bigg( {M\mu_R\over 4\pi} \bigg) 
 \bigg( 2 C_4^{(^1S_0)}(\mu_R)  \: C_0^{(^1S_0)}(\mu_R)+  
 C_2^{(^1S_0)}(\mu_R)^2 \bigg)  \,, \nn \\
\lefteqn{{}^3\!S_1,{}^3\!D_1:\ \ \ \  }  \\
  &&\qquad\quad \beta_0^{({}^3\!S_1)} = \bigg( {M\mu_R\over 4\pi} \bigg) 
  C_0^{({}^3\!S_1)}(\mu_R)^2 \,, 
 \qquad\quad \beta_2^{({}^3\!S_1)} =2 \bigg( {M\mu_R\over 4\pi} \bigg) 
  C_0^{({}^3\!S_1)}(\mu_R)\: C_2^{({}^3\!S_1)}(\mu_R)\,, \nn \\
  && \qquad\quad \beta_2^{({}^3\!S_1- {}^3\!D_1)} =  \bigg( {M\mu_R\over 4\pi} 
 \bigg) C_0^{({}^3\!S_1)}(\mu_R)\:C_2^{({}^3\!S_1- {}^3\!D_1)}(\mu_R)\,, \nn
\end{eqnarray}
in agreement with Refs.~\cite{ksw1,ksw2}.  For $S=0$  states the beta
functions are one loop exact in the sense that the contribution in
Eq.~(\ref{bet02}) comes from the one-loop graphs, with the higher order
graphs giving contributions which cancel.  The reason for this cancellation is
that the only loop corrections are in the bubble chain, and they form a
geometric series.  The sum of bubble graphs is just the chain of irreducible one
loop bubbles for the full (point-like) propagator.  An analogy would be QED, if
the only possible graphs were the two point photon graphs with 
electron loops.  In this case the beta function would also be one-loop exact
because the graphs that are not 1PI do not contribute. 
In general, the beta functions of higher order couplings may have contributions
beyond one-loop in cases where angular momentum mixing is present.

 Expressions for the running coupling constants can 
be derived by summing the counterterms in Eq.~(\ref{ctexpn}) or by solving 
renormalization group equations.  For $s={}^1\!S_0$ or ${}^3\!S_1$ this gives
\begin{eqnarray} \label{rcouplings}
  C_0^{(s)}(\mu_R) &=& {1 \over \frac{1}{C_0^{\rm finite}} - 
	\frac{M\mu_R}{(4\pi)} } \,, \qquad\quad
  C_2^{(s)}(\mu_R) = {C_2^{\rm finite} \over (C_0^{\rm finite})^2 } {1
      \over \bigg[ \frac1{C_0^{\rm finite}} - \frac{M\mu_R}{4\pi}  \bigg]^2 }\,. 
\end{eqnarray}
where $C_0^{\rm finite}$ and $C_2^{\rm finite}$ are constants which can be
determined by specifying boundary conditions.  Since the theory should be
good for arbitrarily small momenta, one possibility is to demand that the
amplitude reproduces the effective range expansion, $p \cot{(\delta)} = -1/a + 
\frac12 r_0 p^2 + {\cal O}(p^4)$.  In Refs.~\cite{ksw1,ksw2} this matching was 
done at $\mu_R=0$ giving $C_0^{\rm finite} = \frac{4\pi a}{M}$, $C_2^{\rm finite} =
\frac{4\pi a}{M}\: \frac{a\, r_0}{2}$, etc.  We could equally well have chosen a
different matching point, and obtained the same results.
For $\mu_R\sim Q$, the running couplings in Eq.~(\ref{rcouplings}) have the 
scaling in Eq.~(\ref{pc}).  Written in terms of renormalized couplings the 
amplitude in the ${}^1\!S_0$ or ${}^3\!S_1$ channels is \cite{ksw2}
\begin{eqnarray}
  {\cal A}
 &=& -\frac{4\pi}{M} \left[ {1\over \frac{4\pi}{MC_0(\mu_R)}+\mu_R+i p} + 
  \frac{4\pi}{M} \frac{C_2(\mu_R)}{C_0(\mu_R)^2} \frac{p^2}{(\frac{4\pi}
  {MC_0(\mu_R)}+\mu_R+i p)^2} + {\cal O}(Q) \right]  \,,   \label{rAmp}
\end{eqnarray}
and satisfies Eq.~(\ref{rc2m}).  The amplitude ${\cal A}$ is $\mu_R$
independent. It is interesting to note that we 
can choose a  renormalization point where all loop corrections vanish giving
\begin{eqnarray}
  {\cal A}^s &=& \sum_{m=0}^\infty {\cal A}^s_{2m} = -\sum_{m=1}^\infty \, 
     C_{2m}^{(s)}(\mu_R=-ip)\,p^{2m} \nn\\
  &=& - \frac{4\pi}{M} \frac1{1/a +ip} - \frac{4\pi}{M}
    \left(\frac1{1/a +ip}\right)^2\, \frac{r_0}2 \: p^2 + \ldots \,.  \label{Atrick}
\end{eqnarray}
The amplitude exactly reproduces the effective range expansion by
construction.   From Eq.~(\ref{Atrick}) the range of the effective field theory can
be estimated as $\Lambda \sim 2/r_0 \sim m_\pi$ as expected.

It is possible to choose the boundary condition for $C_0(\mu_R)$ to change the
location of the pole that appears at each order in the expansion.   For  processes
involving the deuteron \cite{ksw3,metal} a more natural boundary condition is to
choose the pole to appear at $-ip=\gamma=\sqrt{ME_d}$, so $C_0^{\rm
finite}=4\pi/(M\gamma)$.  To recover the effective range expansion,
$C_0(\mu)$ is divided into non-perturbative and perturbative parts\cite{ms0},
$C_0(\mu_R)=C_0^{np}(\mu_R) + C_0^{p}(\mu_R)$, where $C_0^{np}(\mu_R)\sim
1/Q$ and $C_0^{p}(\mu_R)\sim Q^0$.   In this case the amplitude becomes
\begin{eqnarray}
  {\cal A}^s &=&  - \frac{4\pi}{M} \bigg[ \frac1{\gamma +ip} + \frac{4\pi}{M}
   \frac{C_0^{p}(\mu_R)}{(C_0^{np}(\mu_R))^2} \frac1{(\gamma +ip)^2} +
    \frac{4\pi}{M} \frac{C_2(\mu_R)}{(C_0^{np}(\mu_R))^2} \frac1{(\gamma+ip)}
    \: p^2 \bigg]  \,, \label{Ad}
\end{eqnarray}
where the first term is order $1/Q$, and the second and third terms are order
$Q^0$.  The RGE's are
\begin{eqnarray}
  \mu_R \frac{\partial}{\partial\mu_R} C_0^{np}(\mu_R) &=& \frac{M\mu_R}{4\pi} 
     C_0^{np}(\mu_R)^2  \,,\\
  \mu_R \frac{\partial}{\partial\mu_R} C_0^{p}(\mu_R) &=& 2\, \frac{M\mu_R}{4\pi} 
     C_0^{np}(\mu_R) C_0^p(\mu_R) + {\cal O}(Q) \nn \,.
\end{eqnarray}
These can be derived by substituting $C_0(\mu_R)=C_0^{np}(\mu_R) + 
C_0^{p}(\mu_R)$ into the renormalization group equation for $C_0(\mu_R)$.  They
can also be derived using the counterterm method described above. If we demand 
that the observed scattering length and effective range are reproduced at this 
order then we find 
\begin{eqnarray} \label{C0pbc}
 \frac{4\pi}{M} \frac{C_0^{p}(\mu_R)}{(C_0^{np}(\mu_R))^2} = 
    \gamma-\frac1a \,, \qquad\qquad 
 \frac{4\pi}{M} \frac{C_2(\mu_R)}{(C_0^{np}(\mu_R))^2} = \frac{r_0}2  \,.
\end{eqnarray}
In order for the power counting of $C_0^p(\mu_R)$ to be consistent we must treat 
$\gamma-1/a\sim Q^2$.


\section{Theory with nucleons and pions}

\begin{figure}[t!]  
  \epsfysize=9.0truecm \epsfbox{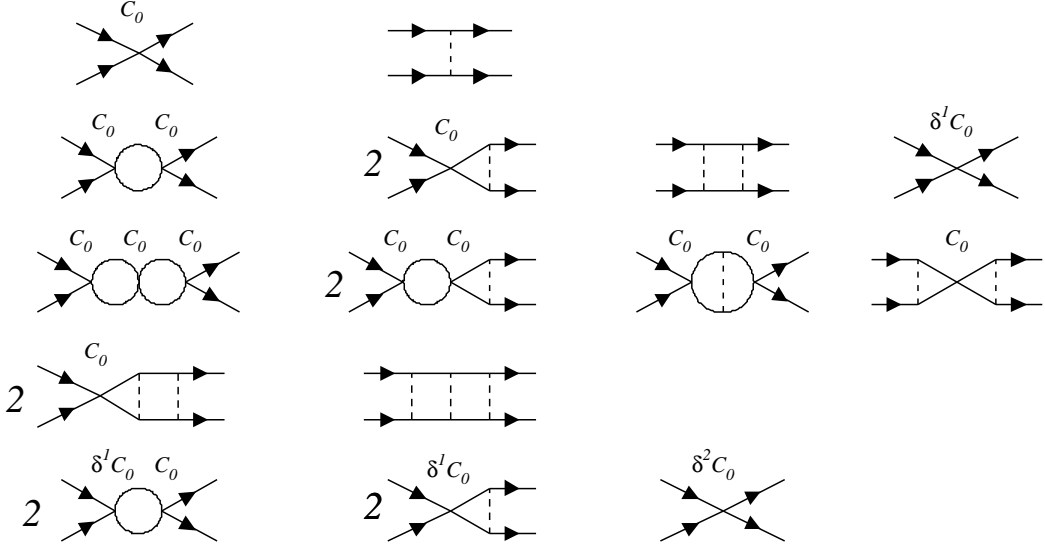}  
{\tighten  \caption[1]{Zero, one, and two-loop graphs with $C_0$ and 
$\delta^nC_0$ vertices and potential pion exchange.  The dashed lines 
denote potential pion propagators.} 
\label{fig_ct0pi} }
\end{figure}  
In this section, we study the renormalization of contact interactions in the
effective field theory with pions.  In the ${}^3\!S_1$ channel, graphs with two or
more consecutive potential pions do not factorize and give poles of the form
$p^n /\epsilon$.  We explicitly compute these poles for two loop pion graphs. 
There are also $m_\pi^2/\epsilon$ poles in both the ${}^1\!S_0$ and ${}^3\!S_1$
channels at order $Q^0$ \cite{ksw1,ksw2}.  Because of these $1/\epsilon$ poles,
pions cannot be summed to all orders in a model independent way.  The finite
counterterms in PDS and OS are different in this theory.   Throughout this
section we will take $m_\pi=0$, since we are only interested in the couplings
$C_{2m}(\mu_R)$.  The $D_2(\mu_R)$ counterterms will be considered in section
V.  We compute the PDS counterterms and beta functions for $C_0(\mu_R)$ and
$C_2(\mu_R)$ to order $Q$.  In PDS, $C_0(\mu_R)$ no longer obeys the $Q$
scaling for $\mu_R \gtrsim 300\,{\rm MeV}$\cite{ksw2}.  This can be fixed by
treating part of the coupling $C_0(\mu_R)$ perturbatively as discussed in
Section VI. The exact expressions for $C_0(\mu_R)$, $C_2(\mu_R)$, and
$C_4(\mu_R)$ are given in the OS scheme and exhibit the correct $Q$ scaling for
all $\mu_R > 1/a $.  Therefore, it is no longer apparent that the power counting
breaks down at $300\,{\rm MeV}$.  The $300\,{\rm MeV}$ scale does appear in the
short distance contribution to the amplitude from pion exchange, however, it can
only be taken as an estimate for the range of the effective field theory once pion
and contact interactions are both included.  In section VI, we will discuss how
experimental data suggests that $\Lambda_\pi\gtrsim 300\,{\rm MeV}$.

To determine how the pions contribute to the beta functions for
$C_{2m}(\mu_R)$, we use the rules in section II.   Some of the pion graphs that
will be needed are shown in Fig.~\ref{fig_ct0pi}.  

\begin{figure}[t!]  
  \centerline{\epsfxsize=15.0truecm \epsfbox{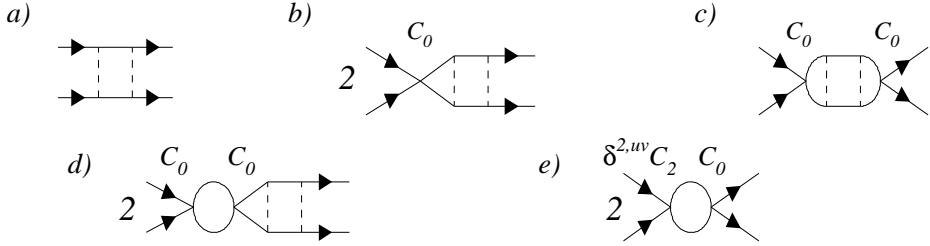}}
{\tighten
 \caption[1]{The basic order $Q$ graphs in the ${}^3S_1$ channel whose loop
  integrals do not factorize even for $m_\pi=0$. }  \label{fig_2ppi} }
\end{figure} 
In both PDS and OS, the first step is to subtract $1/\epsilon$ poles.  For two
nucleons in the ${}^1\!S_0$ channel the spinor indices in Eq.~(\ref{ppip}) are
dotted into $\delta_{\alpha\delta}\, \delta_{\beta\gamma}$.  Therefore the
$m_\pi=0$ piece of pion exchange reduces to a contact interaction and gives
no $1/\epsilon$ poles.  In the ${}^3\!S_1$ channel, graphs with two or more
consecutive pions do not factorize and may have $1/\epsilon$ poles.
Order $Q$ graphs with two consecutive potential pions are shown in the 
first row of Fig.~\ref{fig_2ppi}, and labeled $a)$, $b)$, and $c)$.  We find
\begin{eqnarray} \label{2ppians}
  a)\quad  &=&\ -i\: \frac32\: \bigg(\frac{g_A^2}{2f^2}\bigg)^2 \bigg(
	\frac{-ip\,M}{4\pi} \bigg) \,, \\ 
  b)\quad &=&\ -3 i\: C_0^{\rm finite}\: \bigg(\frac{g_A^2}{2f^2}\bigg)^2 \:\bigg(
	\frac{-ip\,M}{4\pi}\bigg)^2 \left[ \frac1{\epsilon}-2\gamma+\frac{14}{3}
	-4\ln{(2)} + 2\,\ln{\bigg(\frac{\pi\mu^2}{-p^2-i\varepsilon}\bigg)}\right]
	 \,, \nn \\
  c)\quad &=& -3 i\:(C_0^{\rm finite})^2\: \bigg(\frac{g_A^2}{2f^2}\bigg)^2 \: 
	\bigg( \frac{-ip\,M}{4\pi}\bigg)^3 \left[ \frac1{\epsilon}-3\gamma-
	6\ln{2} +\frac{37}{6} + 3\,\ln{\bigg(\frac{\pi\mu^2}{-p^2-
	i\varepsilon}\bigg)} \right] \,.\nn
\end{eqnarray}
Graphs $b)$ and $c)$ have been written with $C_0^{\rm finite}$ vertices to 
emphasize that the uv counterterm which cancels 
their divergent part is independent of $\mu_R$.   The divergence in $b)$ is 
cancelled by a tree level graph with the counterterm
\begin{eqnarray} \label{C2infct}
  \delta^{2,\,\rm uv}C_2 = -\, 6\, C_0^{\rm finite} \bigg( \frac{M g_A^2}
  {8\pi f^2} \bigg)^2 \left[ \frac1{2\epsilon} -\gamma + \ln{(\pi)} +2-2\ln{(2)} 
  \right] \,,
\end{eqnarray}
where the superscript $2$ indicates that the counterterm comes in at two
loops. The extra factor $2-2\ln{(2)}$ is included because this leads to
simpler analytic expressions.  Expanding the $C_0$ bubble graph
(second row, first column of Fig.~\ref{fig_ct0pi}) in $\epsilon$ gives
\begin{eqnarray}  \label{c0expn}
   - \frac{p M}{4\pi} (C_0)^2 \left\{ 1 + \epsilon\bigg[ 2-\gamma-2\ln{(2)}+
     \ln{\bigg( \frac{\pi\mu^2}{-p^2-i\epsilon} \bigg) \bigg]}  \right\}  \,.
\end{eqnarray} 
When graphs with $1/\epsilon$ poles are dressed with $C_0$ bubbles, the
factors of $[2-\gamma-2\ln{2} + \ln{(\pi)}]$ that appear are cancelled by similar
factors from the counterterms.  In fact, $\delta^{2,\,\rm uv}C_2$ is the only uv 
counterterm we need for two potential pion exchange
with $m_\pi=0$.  The $1/{\epsilon}$ pole in $c)$ is nonanalytic since it is
proportional to $p^3$.  When graph $c)$ is added to graphs $d)$ and $e)$ the
poles cancel.  These cancellations continue to occur when more $C_0$
bubbles are added to $b)$ and $c)$.  After including graphs with 
$\delta^{2,\rm uv}C_2$ we find
\begin{eqnarray}  \label{newbc}
  b)+i\delta^{2,\,\rm uv}C_2 \:p^2\quad &=&\ -3 i\: C_0^{\rm finite}\: 
    \bigg(\frac{g_A^2}{2f^2}\bigg)^2 \:\bigg(
    \frac{-ip\,M}{4\pi}\bigg)^2 \left[ \frac23+ 
    2\,\ln{\bigg(\frac{\mu^2}{-p^2-i\varepsilon}\bigg)} \right] \,, \\[5pt]
  c)+\frac12 e) \quad &=& -3 i\:(C_0^{\rm finite})^2\: \bigg(\frac{g_A^2}{2f^2}
     \bigg)^2 \: \bigg( \frac{-ip\,M}{4\pi}\bigg)^3  \left[\frac16 + 2 \,
     \ln{\bigg(\frac{\mu^2}{-p^2-i\varepsilon}\bigg)}\right] \ \,. \nn
\end{eqnarray}
Note that for $\mu \sim p$ there are no large numerical factors from these
graphs.  

In the $^3S_1$ channel, potential pion graphs without contact interactions 
also have $p^2/\epsilon$ poles.  The two loop graph with three 
potential pions (fourth row, second column in Fig.~\ref{fig_ct0pi}) is equal to
\begin{eqnarray}
   \frac{4\pi i }{M} \bigg( \frac{M g_A^2}{8\pi f^2} \bigg)^3 \: p^2\ \bigg[ 
 	\frac3{\epsilon} + \ldots \bigg] \,.
\end{eqnarray}
In the $Q$ power counting, this graph is order $Q^2$ and will not be
considered here.  Because of these $1/\epsilon$ poles it is not possible to sum
pion ladder graphs to all orders.  Now that the ultraviolet divergences have been 
removed from graphs $b)$ and $c)$, the finite subtractions can be performed.  

\subsection{PDS}

For PDS in the $^1S_0$ channel, we can compute the effect of potential pions
on the $C_{2m}(\mu_R)$ counterterms to all orders in $Q$ (neglecting
relativistic corrections).  For $C_0(\mu_R)$, the relevant zero, one, and two 
loop graphs are  shown in Fig.~\ref{fig_ct0pi}.  The $C_0(\mu_R)$  and 
$C_2(\mu_R)$ counterterms are
\begin{eqnarray}
 && \delta^nC_0^{({}^1\!S_0)}(\mu_R) =(-1)^{n+1} \bigg( \frac{M\mu_R}{4\pi}
  \bigg)^n  \bigg[ C_0(\mu_R)+\frac{g_A^2}{2f^2} \bigg]^{n+1} \,, \nn \\
 && \delta^nC_2^{({}^1\!S_0)}(\mu_R) = (-1)^{n+1} (n+1) \bigg( \frac{M\mu_R}
  {4\pi} \bigg)^n  \bigg[ C_0(\mu_R)+\frac{g_A^2}{2f^2} \bigg]^{n} 
   C_2(\mu_R) \,. \label{1s0ct} 
\end{eqnarray}

The PDS counterterms in the ${}^3\!S_1$ channel will only be computed to
order $Q$ since the loop graphs with consecutive pions do not factorize.  For
this case it is essential to use the counterterms to carry out the PDS
renormalization program.  To define $C_0(\mu_R)$ at order $Q$, we set up the
finite subtractions as in Fig.~\ref{fig_ct0pi}, but leave out all graphs with more
than two potential pions since they are ${\cal O}(Q^2)$ (we also neglect
relativistic corrections that are order $Q$ but come with an additional
$1/M^2$).  Note that in $d=3$ only the overall divergence ($\propto 1/(d-3)^n$
for $n$ loops) is needed since loops with counterterms will cancel the
sub-divergences.  Evaluating  the graphs in Fig.~\ref{fig_2ppi} with $d=3$ and
then continuing back to $d=4$ gives
\begin{eqnarray}
  a)&=& -9\, i\, \bigg(\frac{g_A^2}{2f^2}\bigg)^2 \bigg(\frac{\mu_R\,M}{4\pi} \bigg),
  \qquad
  b) = -12\, i\, C_0(\mu_R) \bigg(\frac{g_A^2}{2f^2}\bigg)^2 
	\bigg(\frac{\mu_R\,M}{4\pi} \bigg)^2  \,, \nn\\
  c) &=& -5\, i\, C_0(\mu_R)^2 \bigg(\frac{g_A^2}{2f^2}\bigg)^2 
	\bigg(\frac{\mu_R\,M}{4\pi} \bigg)^3 \,.
\end{eqnarray}
Using these values we find
\begin{eqnarray}   \label{3s1ct0} 
 \delta^1C_0^{(^3S_1)} &=&  \bigg( \frac{M\mu_R}{4\pi}\bigg) \bigg[
   C_0(\mu_R)^{2}+2 C_0(\mu_R) \frac{g_A^2}{2f^2}  + 9\, 
    \Big(\frac{g_A^2}{2f^2}\Big)^2   \bigg] \,, \\
 \delta^nC_0^{({}^3\!S_1)} &=& (-1)^{n+1} \bigg( \frac{M\mu_R}{4\pi}\bigg)^n
   \bigg[ C_0(\mu_R)^{n+1}+(n+1) C_0(\mu_R)^n \frac{g_A^2}{2f^2} 
  \nn\\
& &\qquad\qquad+\frac12 (n+1)
   (n+4) C_0(\mu_R)^{n-1} \Big(\frac{g_A^2}{2f^2}\Big)^2
   \bigg] \,. \qquad\mbox{for $n\ge2$}  \nn
\end{eqnarray}
Note that for graphs with two consecutive potential pions, the $\mu_R$ 
dependence does not come in the linear combination $\mu_R+ip$.  For 
instance, adding the PDS counterterm to graph a) in Fig.~\ref{fig_2ppi} gives
the linear combination $3 ip/2 + 9\mu_R$.

In PDS, like in $\overline{\rm MS}$, the renormalized coupling $C_2(\mu_R)$ will
depend on $\ln(\mu_R^2/\mu_0^2)$ in such a way that the $\ln(\mu_R^2)$
dependence in the amplitudes in Eq.~(\ref{2ppians}) is cancelled.  Here $\mu_0$ is
an arbitrary scale expected to be of order $\Lambda_\pi$.  At order $Q$ we find 
\begin{eqnarray} \label{3s1ct2}
  \delta^1C_2^{(^3S_1)}(\mu_R) &=&  2\bigg( \frac{M\mu_R}{4\pi} \bigg)
      \bigg[ C_0(\mu_R)+ \frac{g_A^2}{2f^2} \bigg] C_2(\mu_R)\,, \\
  \delta^nC_2^{(^3S_1)}(\mu_R) &=& (-1)^{n+1} \left\{ (n+1) \bigg( 
     \frac{M\mu_R}{4\pi} \bigg)^n  \bigg[ C_0(\mu_R)^n+ n \frac{g_A^2}{2f^2} 
    C_0(\mu_R)^{n-1} \bigg] C_2(\mu_R) \right. \nn  \\
 && \qquad\qquad\quad \left. + 6 \bigg( \frac{M\mu_R}{4\pi}
     \bigg)^{n-2} C_0(\mu_R)^{n-1} \bigg( \frac{M g_A^2}{8\pi f^2} \bigg)^2  
     \ln{\Big(\frac{\mu_R^2}{\mu_0^2}\Big)} \right\}
\qquad\mbox{for $n\ge2$} \nn \,.
\end{eqnarray}
Note that the part of $\delta^nC_2(\mu_R)$ proportional to 
$\ln{(\mu_R^2/\mu_0^2)}$ has a coefficient that sums up to $C_0^{\rm finite}$ at
this order.  From Eqs.~(\ref{1s0ct}), (\ref{3s1ct0}), and (\ref{3s1ct2}) we find
\begin{eqnarray}  \label{PDSbeta}
  \beta_0^{({}^1\!S_0)} &=&  \frac{M\mu_R}{4\pi}\: \bigg[C_0(\mu_R)+
   \frac{g_A^2}{2f^2}  \bigg]^2  \,, \qquad
  \beta_2^{({}^1\!S_0)} =  2\: \frac{M\mu_R}{4\pi}\:  \bigg[C_0(\mu_R)+
   \frac{g_A^2}{2f^2}  \bigg] C_2(\mu_R) \,, \\[5pt]
  \beta_0^{({}^3\!S_1)} &=&  \frac{M\mu_R}{4\pi} \left\{ C_0^2+
   2 \frac{g_A^2}{2f^2} C_0 + \bigg[9+4\bigg(\frac{\mu_R M C_0}{4\pi}\bigg) 
   +2\bigg(\frac{\mu_R M C_0}{4\pi}\bigg)^2 \bigg]
   \Big(\frac{g_A^2}{2f^2}\Big)^2 \right\} + {\cal O}(Q^2) \,, \nn \\[5pt]
  \beta_2^{({}^3\!S_1)} &=& 2\: \frac{M\mu_R}{4\pi}\: \bigg[C_0(\mu_R)+
   \frac{g_A^2}{2f^2}  \bigg] C_2(\mu_R) -12 \bigg(\frac{M g_A^2}{8\pi f^2}
   \bigg)^2 C_0(\mu_R) \bigg[ 1+ \mu_R \frac{M}{4\pi} C_0(\mu_R) \bigg]
  +{\cal O}(Q^0) \nn \,.
\end{eqnarray}
Note that in the ${}^1\!S_0$ channel all contributions to the beta functions
beyond one-loop cancel, leaving them one-loop exact.  In  Ref.~\cite{ksw2},
the last two terms in $\beta_0^{(^3S_1)}$ are absent, but should be included in
the complete order $Q$ calculation.  Dimensional analysis implies that the
${}^3\!S_1$ beta functions can have corrections at all higher orders in Q, since
there is nothing to prevent the dimensionless factor $(\mu_R\,g_A^2 \,M)/(8\pi
f^2) \sim Q$ from appearing.  In Ref.~\cite{sf2}, expressions for the beta
functions are derived by demanding that $\partial{\cal A}/\partial\mu_R =0$, but
these are not the PDS beta functions.  Since in all renormalization schemes
$\partial{\cal A}/ \partial\mu_R=0$, this condition is not sufficient to fix the
renormalization scheme uniquely.  As discussed in Ref.~\cite{ms0}, the large
$\mu_R$ behavior of $C_0^{(^3S_1)}(\mu_R)$ is unknown because of the 
higher order corrections.

\subsection{OS}

In the OS scheme, there is no such ambiguity since at a given order in $Q$ the
running of all the coupling constants that enter at that order are known
exactly.  The coupling $C_0^{(s)}(\mu_R)$ has contributions only from the
nucleon graphs discussed in section II and therefore has the same beta
function.  For $C_2^{(s)}(\mu_R)$, the order $Q^0$ graphs in ${\cal A}_2$
include the nucleon graphs from section II, as well as the graphs with one
potential pion and any number of $C_0$ vertices.   At tree level we add a finite
counterterm to cancel the $m_\pi=0$ part of the tree level pion interaction at
$p=i\mu_R$
\begin{eqnarray}
  \delta^0 C_2^{(s)}(\mu_R) = -\ \frac{g_A^2}{2f^2}\ \frac1{\mu_R^2} \,.
\end{eqnarray}
This counterterm is order $Q^{-2}$ like $C_2(\mu_R)$ itself.  Since all the 
graphs in ${\cal A}_2$ factorize the higher loop counterterms are the same as 
in the theory without pions, so $\delta^n C_2$ for $n\ge 1$ are given in 
Eq.~(\ref{ct024}).  The exact beta function is then
\begin{eqnarray}
  \beta_2^{(s)} = 2\: \frac{M\mu_R}{4\pi}\: C_0(\mu_R) C_2(\mu)+
    2\:\frac{g_A^2}{2f^2} \bigg( 1+\frac{M\mu_R}{4\pi}C_0(\mu_R)
  \bigg)^2\: \frac1{{\mu_R}^2}  \,,
\end{eqnarray}
Note that the finite $\ln(\mu^2/(-p^2-i\epsilon))$ terms in Eq.~(\ref{2ppians}) 
are order $Q$ and in the OS scheme do not affect the running of 
$C_2(\mu_R)$, but rather $C_4(\mu_R)$.  In terms of the finite 
constants $C_0^{\rm finite}$ and $C_2^{\rm finite}$ we have solutions 
\begin{eqnarray} \label{OSc0c2}
  C_0^{(s)}(\mu_R) &=& {1 \over \frac{1}{C_0^{\rm finite}} - 
	\frac{M\mu_R}{(4\pi)} } \,, \qquad\quad
  C_2^{(s)}(\mu_R) = {C_2^{\rm finite} - \frac{g_A^2}{2f^2\mu_R^2} 
      \over \left[ 1 - \mu_R \frac{C_0^{\rm finite} M}{4\pi}  \right]^2 }\,. 
\end{eqnarray}
Although it may seem that the piece of $C_2^{(s)}(\mu_R)$ that goes as
$1/\mu_R^4$ will spoil the power counting for low momentum, in fact, the
$1/\mu_R^2$ part dominates entirely until $\mu_R\sim 1/a$, since $C_0^{\rm
finite} \sim a$, $C_2^{\rm finite}\sim a^2$.    Written in terms of
renormalized couplings the $m_\pi=0$ part of the next-to-leading order
OS amplitude is
\begin{eqnarray}
 { - C_2(\mu_R) \:p^2 \over \bigg[1 + (\mu_R+ip)\frac{M C_0(\mu_R)}
   {4\pi} \bigg]^2 } \ -\  \frac{g_A^2}{2f^2}\: {\mu_R^2 + p^2 \over \mu_R^2} \:
    { \bigg[ 1+ \mu_R \frac{C_0(\mu_R) M}{4\pi}  \bigg]^2 \over \bigg[ 1+
   (\mu_R+ip) \frac{C_0(\mu_R) M}{4\pi} \bigg]^2 } \,,
\end{eqnarray}
which is order $Q^0$ as desired.  

One might still ask if the problem with the $300\,{\rm MeV}$ scale will reappear
in higher order coefficients.  To check that this is not the case we compute
the running of the coupling $C_4(\mu_R)$ in the OS scheme.  
The  easiest way to compute this running coupling constant is to
compute the order $Q$ amplitude in terms of the finite couplings,
$C_{2m}^{\rm finite}$, and then demand that the amplitude satisfies the
renormalization condition in Eq.~(\ref{rc2m}).  The graphs we need to compute 
include those with
\begin{eqnarray} \begin{array}{rl}
    i)\ &\mbox{ one $C_4$ and any number of $C_0$'s}  \,, \\
    ii)\  &\mbox{ two $C_2$'s and any number of $C_0$'s}  \,, \\
    iii)\  &\mbox{ one $C_2$, one potential pion and any number of $C_0$'s}\,, \\
    iv)\  &\mbox{ two potential pions and any number of $C_0$'s}  \,.
\end{array}
\end{eqnarray}
Computing these graphs in terms of the finite couplings and then demanding
that the amplitudes satisfy the renormalization condition gives the OS 
couplings
\begin{eqnarray}  \label{C4sln}
  C_4^{({}^1\!S_0)}(\mu_R) &=& {C_4^{\rm finite}\over\Big[1-\mu_R\frac{M}{4\pi}
     C_0^{\rm finite}\Big]^2 } + \frac{\mu_R M}{4\pi} { \Big[C_2^{\rm finite}  -
     g_A^2/(2f^2)\, \frac1{{\mu_R}^2} \Big]^2 \over \Big[1-\mu_R\frac{M}{4\pi} 
     C_0^{\rm finite}\Big]^3 } \,, \\
  C_4^{({}^3\!S_1)}(\mu_R) &=&{C_4^{\rm finite}\over\Big[1-\mu_R\frac{M}{4\pi}
     C_0^{\rm finite}\Big]^2 } + \frac{\mu_R M}{4\pi} { \Big[C_2^{\rm finite}  -
     g_A^2/(2f^2)\, \frac1{{\mu_R}^2} \Big]^2 \over \Big[1-\mu_R\frac{M}{4\pi} 
     C_0^{\rm finite}\Big]^3 }  \nn\\
  + \frac12 \bigg( \frac{g_A^2}{2f^2} \bigg)^2 \!\!\!&&\!\!\!\!
     \frac{M}{4\pi}\, \frac1{{\mu_R}^3}\, {\bigg[ 1-2\mu_R \frac{M}{4\pi} 
      C_0^{\rm finite} \bigg] \over \bigg[1-\mu_R  \frac{M}{4\pi} C_0^{\rm finite} 
      \bigg]^2 } -  
  6\,\bigg( \frac{M g_A^2}{8\pi f^2}\bigg)^2 \frac{ C_0^{\rm finite} 
  \ln{( \mu_R^2 /\mu^2 )}} {\mu_R^2 \bigg[1-\mu_R \frac{M}{4\pi} 
   C_0^{\rm finite}\bigg]  }  \,, \nn
\end{eqnarray}
where here $\mu$ is an unknown scale expected to be of order $\Lambda_\pi$. 
Again the pion contributions do not spoil the $\mu_R$ scaling behavior,
since they are suppressed by factors of the large scattering length.  Note that
at order $Q$ the PDS coupling $C_4(\mu_R)$ \cite{ksw2} is the $g_A\to 0$ limit of 
Eq.~(\ref{C4sln}). 

In this section, expressions for the renormalized couplings $C_0(\mu_R)$,
$C_2(\mu_R)$, and $C_4(\mu_R)$ were derived in the PDS and OS schemes
working to order $Q$.   For the $^3S_1$ channel, we have shown that
$C_0(\mu_R)$ has corrections at all orders in $Q$ in PDS.  Unlike PDS, the OS
couplings $C_{2m}(\mu_R)$ can be computed exactly because they only have
contributions at one order in $Q$.  The OS couplings exhibit the correct
$\mu_R$ scaling for all $\mu_R>1/a$.


\section{The coupling $D_2(\mu_R)$}

In this section, the OS and PDS counterterms for $D_2(\mu_R)$ are computed.
To define $D_2(\mu_R)$ in the OS scheme, we take
\begin{eqnarray}
   &&  i\,A^s(D_2) \Big|_{p=i\mu_R,} =  -i\, D_2^{(s)}(\mu_R)\, m_\pi^2 \,,
\end{eqnarray}
where ${\cal A}^s(D_2)$ contains terms in the amplitude that are analytic in
$m_\pi^2$ and proportional to $m_\pi^2$.  Only terms that are analytic in
$m_\pi^2$ are kept because it is unnatural to put  long-distance nonanalytic
contributions that come from pion exchange into the definition of the short
distance coupling \cite{sf2}.  For example, one potential pion exchange gives a
$m_\pi^2/p^2 \: \ln(1+4p^2/m_\pi^2)$ term. Including this in $A^s(D_2)$ would give
$D_2(\mu_R)$ both a branch cut at $\mu_R=m_\pi/2$ as well as explicit
dependence on the scale $m_\pi$.  In the OS scheme, $D_2(\mu_R)$ will be
calculated as follows.  First $m_\pi^2/\epsilon$ poles are subtracted.  The finite
counterterms are then determined by including graphs with a single $D_2(\mu_R)$
or potential pion and any number of $C_0(\mu_R)$ vertices in ${\cal A}^s(D_2)$. 
Contributions from these graphs that are non-analytic in $m_\pi^2$ are dropped.

There is a $m_\pi^2/\epsilon$ pole in the ${\cal O}(Q^0)$ graph in the third row
and third column of Fig.~\ref{fig_ct0pi} \cite{ksw1,ksw2}, so we have a
counterterm
\begin{eqnarray}
  \delta^{2,{\rm uv}}D_2 = - i \bigg(\frac{MC_0^{\rm finite}}{4\pi}\bigg)^2 \:
   \frac{g_A^2}{4f^2} \:  \left[ \frac1{2\epsilon} -\gamma + \log{(\pi})  \right] \,.
\end{eqnarray}
Note that when this counterterm is dressed with $C_0$ bubbles the  
extra factors of $2-\ln 2$ from Eq.~(\ref{c0expn}) will cancel without the need for
an additional finite term in $\delta^{2,{\rm uv}}D_2$.  After subtracting this 
counterterm the value of the two-loop graph is 
\begin{eqnarray} \label{tlp}
    i \bigg(\frac{MC_0^{\rm finite}}{4\pi}\bigg)^2 \:
   \frac{g_A^2}{2f^2} \:  \left[ -(ip)^2 + \frac{m_\pi^2}{2} +\frac{m_\pi^2}{2}
    \ln{\bigg(\frac{\mu^2}{m_\pi^2}\bigg)}- m_\pi^2 
    \ln{\bigg(1-\frac{2ip}{m_\pi}\bigg)} \right] \,.
\end{eqnarray}
For PDS we set $\mu=\mu_R$ and then find finite counterterms 
\begin{eqnarray} \label{3s1ctd2}
  \delta^1D_2(\mu_R) &=&  2\bigg( \frac{M\mu_R}{4\pi} \bigg)
    C_0(\mu_R) D_2(\mu_R) \,, \\
  \delta^nD_2(\mu_R) &=& (-1)^{n+1} \bigg[ (n+1) \bigg( \frac{M\mu_R}{4\pi}
     \bigg)^n  C_0(\mu_R)^n D_2(\mu_R) \\
  && \qquad\quad\quad - \frac{(n-1)}2 \bigg( \frac{M\mu_R}{4\pi}
     \bigg)^{n-2} C_0(\mu_R)^{n} \bigg( \frac{M}{4\pi}\bigg)^2 
     \frac{g_A^2}{2 f^2}   \ln{\Big(\frac{\mu_R^2}{\mu_0^2}\Big)} \bigg]
\qquad\mbox{for $n\ge2$} \nn \,.
\end{eqnarray}
Here $\mu_0$ is an unknown scale expected to be of order $\Lambda_\pi$.

In the OS scheme, the $\delta^1D_2(\mu_R)$ counterterm is the same as in PDS. 
In dimensional regularization logarithms of the form $\ln(\mu^2/m_\pi^2)$ will
appear in loop graphs.  To make the $\mu^2$ dependent part analytic in 
$m_\pi^2$ we write
\begin{eqnarray}
  \ln\left(\frac{\mu^2}{m_\pi^2}\right) = \ln\left(\frac{\mu^2}{\mu_R^2}\right) + 
	\ln\left(\frac{\mu_R^2}{m_\pi^2}\right) \,,
\end{eqnarray}
and then subtract the $\ln(\mu^2/\mu_R^2)$ term with the counterterms.  This
will give $D_2(\mu_R)$ a $\mu_R$ dependence which cancels the 
$\ln(\mu_R^2/m_\pi^2)$ in the amplitude. In the OS scheme, the $m_\pi^2/2$ in
Eq.~(\ref{tlp}) gets subtracted along with the logarithm.  We  find
\begin{eqnarray} \label{3s1ctd2os}
  \delta^nD_2(\mu_R) &=& (-1)^{n+1} \Bigg\{ (n+1) \bigg( \frac{M\mu_R}{4\pi}
     \bigg)^n  C_0(\mu_R)^n D_2(\mu_R) \nn\\
  && \quad - \frac{(n-1)}2 \bigg( \frac{M\mu_R}{4\pi}
     \bigg)^{n-2} C_0(\mu_R)^{n} \bigg( \frac{M}{4\pi}\bigg)^2 
     \frac{g_A^2}{2 f^2} \bigg[-1+  \ln{\Big(\frac{\mu_R^2}{\mu^2}\Big)} 
     \bigg] \ \Bigg\}
\qquad\mbox{for $n\ge2$} \nn \,.
\end{eqnarray}
Summing the counterterms the solutions for $D_2(\mu_R)$ are then
\begin{eqnarray} \label{D2sln}
 {D_2^{(s)}(\mu_R)\over C_0^{(s)}(\mu_R)^2} &=& { D_2^{\rm finite} \over 
   (C_0^{\rm finite})^2}  +\frac{M}{8\pi} \bigg(\frac{M g_A^2}{8\pi f^2} \bigg) 
   \ln\bigg({\mu_R^2 \over \mu_0^2}\bigg)\qquad\qquad\quad\mbox{in PDS} \,,\\
 {D_2^{(s)}(\mu_R)\over C_0^{(s)}(\mu_R)^2} &=& { D_2^{\rm finite} \over 
   (C_0^{\rm finite})^2}  +\frac{M}{8\pi} \bigg(\frac{M g_A^2}{8\pi f^2} \bigg) 
   \bigg[-1+  \ln\bigg({\mu_R^2 \over \mu^2}\bigg)  \bigg]
   \qquad \mbox{in OS} \,, \nn
\end{eqnarray}
which can be written as
\begin{eqnarray}
{D_2^{(s)}(\mu_R)\over C_0^{(s)}(\mu_R)^2}  &=& \frac{M}{8\pi} 
  \bigg(\frac{M g_A^2} {8\pi f^2} \bigg)  \ln\bigg({\mu_R^2  \over \tilde\mu^2}
  \bigg) \,, \\[5pt] \mbox{  where    } 
  && \tilde\mu^2 = \mu_0^2 \, \exp{\bigg(\frac{-64\pi^2 f^2  D_2^{\rm finite}}
  {M^2  g_A^2(C_0^{\rm finite})^2} \bigg)}   \qquad\mbox{in PDS} \,, \\
  && \tilde\mu^2 = \mu^2 \, \exp{\bigg(1-\frac{64\pi^2 f^2  D_2^{\rm finite}}
  {M^2  g_A^2(C_0^{\rm finite})^2} \bigg)}   \quad\mbox{in OS} \,.  
   \label{D2ln} 
\end{eqnarray} 
The parameter $\tilde \mu$ must be determined by fitting to data.  
With $m_\pi\sim Q\sim \mu_R$, $D_2(\mu_R)m_\pi^2 \sim Q^0$ in both OS and
PDS, implying that $D_2(\mu_R)$ should be treated perturbatively.


\section{Schemes and Amplitudes} 

In this section, the amplitudes in the $^1S_0$ and $^3S_1$ channels are presented
to order $Q^0$, both in PDS \cite{ksw1,ksw2} and  OS. Fits to the $^1S_0$ and
$^3S_1$ phase shift data are done in both schemes for different values of $\mu_R$.
As pointed out in Ref.~\cite{ms0}, one  has the freedom to split $C_0(\mu_R)$ into
perturbative and nonperturbative pieces: $C_0(\mu_R)=C_0^{np}(\mu_R) +
C_0^p(\mu_R)$, where $C_0^{np}(\mu_R) \sim Q^{-1}$ and $C_0^p(\mu_R) \sim
Q^0$.  This division is necessary in PDS in order to obtain $\mu_R$ independent
amplitudes at each order.  Furthermore, $C_0^{np}(\mu_R)\sim 1/\mu_R$ so the
coefficients scale in a manner consistent with the power counting for all
$\mu_R>1/a$.  For convenience we will drop the superscript $np$ in what follows. 
Some issues that arise in matching the pion theory onto the effective range
expansion are also discussed.

First, we give the nucleon-nucleon scattering amplitudes in the PDS and OS
schemes.  In PDS, the amplitudes were calculated to order $Q^0$ in
Refs.~\cite{ksw1,ksw2}. At this order, amplitudes in the  ${{}^1\!S_0}$ and
${{}^3\!S_1}$ channels have the same functional form,
\begin{eqnarray}
  A &=& A^{(-1)} + A^{(0,a)} + A^{(0,b)} + {\cal O}(Q^1) \,.
\end{eqnarray} 
In both OS and PDS we have
\begin{eqnarray}  \label{PDSA}
 {A^{(-1)}} &=& -\frac{4\pi}{M}\: \frac{1} {\frac{4\pi}{M C_0(\mu_R)} +
	\mu_R + ip } \,,  \\[10pt]  
 \frac{A^{(0,a)}}{\Big[{A^{(-1)}}\Big]^2} &=&  \frac{ g_A^2 m_\pi^2 }{2 f^2} \bigg( \frac{M}{4\pi} \bigg)^2 \left\{ 
  \frac12 \ln{\bigg({ \mu_R^2\over m_\pi^2} \bigg)} - \bigg(\frac{4\pi}
  {MC_0(\mu_R)} + \mu_R\bigg) \frac{1}{p} \tan^{-1}\bigg(\frac{2p}{m_\pi}\bigg)
  \right. \nn \\[5pt]
&&\left.\  + \bigg[ \bigg(\frac{4\pi}{MC_0(\mu_R)} 
  +\mu_R\bigg)^2-p^2 \bigg] \frac1{4p^2} \ln{\bigg(1+ \frac{4p^2}{m_\pi^2} 
  \bigg)} \right\}  - \frac{D_2(\mu_R)\: m_\pi^2 }{C_0(\mu_R)^2} \nn \,.
\end{eqnarray}
The remaining part of the order $Q^0$ PDS amplitude is
\begin{eqnarray}  
 \frac{A^{(0,b)}}{\Big[{A^{(-1)}}\Big]^2} &=& -\frac{ C_2(\mu_R)\: p^2}
  {C_0(\mu_R)^2} + \frac12 \: \frac{ g_A^2 m_\pi^2 }{2 f^2} \bigg(\frac{M}{4\pi} 
  \bigg)^2 -\: \frac1{C_0(\mu_R)^2} \bigg[\frac{g_A^2}{2 f^2} + C_0^p(\mu_R) 
  \bigg] \,.
\end{eqnarray}
Note that since we have made a different finite subtraction than KSW the second
term has a prefactor of $1/2$, rather than a $1$ as in Ref.~\cite{ksw2}.  In the OS 
scheme 
\begin{eqnarray}  \label{OSA}
\frac{A^{(0,b)}}{\Big[{A^{(-1)}}\Big]^2} &=& 
  -\frac{ C_2(\mu_R)\: p^2} {C_0(\mu_R)^2}  -\frac{g_A^2}
  {2 f^2} \bigg(1+\frac{p^2}{\mu_R^2}\bigg) \bigg( \frac1{C_0(\mu_R)} + 
  \frac{M\mu_R}{4\pi} \bigg)^2 -\: \frac{C_0^p(\mu_R)}{C_0(\mu_R)^2}
   \label{Amp4} \,.
\end{eqnarray}
In appendix B, we give relations between the OS and PDS couplings that appear
at this order.  Using these equations it is easy to verify that the amplitudes are 
equivalent in the two schemes.

In the PDS scheme, there are order $Q^0$ contributions to $\beta_0$ (c.f.,
Eq.~(\ref{PDSbeta})).  If the order $Q^0$ contributions are separated from the 
order $1/Q$ pieces, the beta function for $C_0^p(\mu_R)$ is
\begin{eqnarray}  \label{C0pb}
  \mu_R \frac{\partial C_0^p(\mu_R)}{\partial \mu_R}  =  2\,\frac{M\mu_R}{4\pi} 
    C_0(\mu_R) \bigg[ C_0^p(\mu_R) + \frac{g_A^2}{2f^2} \bigg] +{\cal O}(Q)\,.
\end{eqnarray}
This equation has the solution
\begin{eqnarray} \label{C0ps}
    \frac{C_0^p(\mu_R)}{C_0(\mu_R)^2} = \frac{M}{4\pi} K 
	-\frac{g_A^2}{2f^2} \frac1{C_0(\mu_R)^2} \,,
\end{eqnarray}
where $K$ is a constant which must be order $Q^2$ for $C_0^p(\mu_R)\sim 
Q^0$.  (Recall, from Eq.~(\ref{C0pbc}) that $K=\gamma-1/a \lesssim 1/a$ in the 
pure nucleon theory.) Including $C_0^p(\mu_R)$ makes the PDS amplitudes
explicitly $\mu_R$ independent.  In performing fits to the data the constant $K$
and coupling $D_2(\mu_R)$ cannot be determined independently.  In what
follows we will drop $K$ and simply remember that the values of $D_2(\mu_R)$
extracted from the fits may differ from the renormalized coupling in the
Lagrangian.  In PDS, if $C_0^p(\mu_R)$ is omitted from our expressions then 
$D_2(\mu_R)$ does not follow the renormalization group equation, as we will see 
below.

In the OS scheme, the constant $g_A^2/(2f^2)$ in Eqs.(\ref{C0pb}) and (\ref{C0ps})
is not present, so $C_0^p(\mu_R)/C_0(\mu_R)^2$ is $\mu_R$ independent.   The
OS scheme amplitudes $A^{(-1)}$ and $A^{(0)}$ are therefore $\mu_R$
independent without $C_0^p(\mu_R)$ as can be seen by examining
Eqs.~(\ref{OSc0c2}), and (\ref{D2sln}).  In OS the constant $K$ will also be
absorbed into $D_2(\mu_R)$.
 
\begin{figure}[!t]  
  \centerline{\epsfxsize=8.0truecm \epsfbox{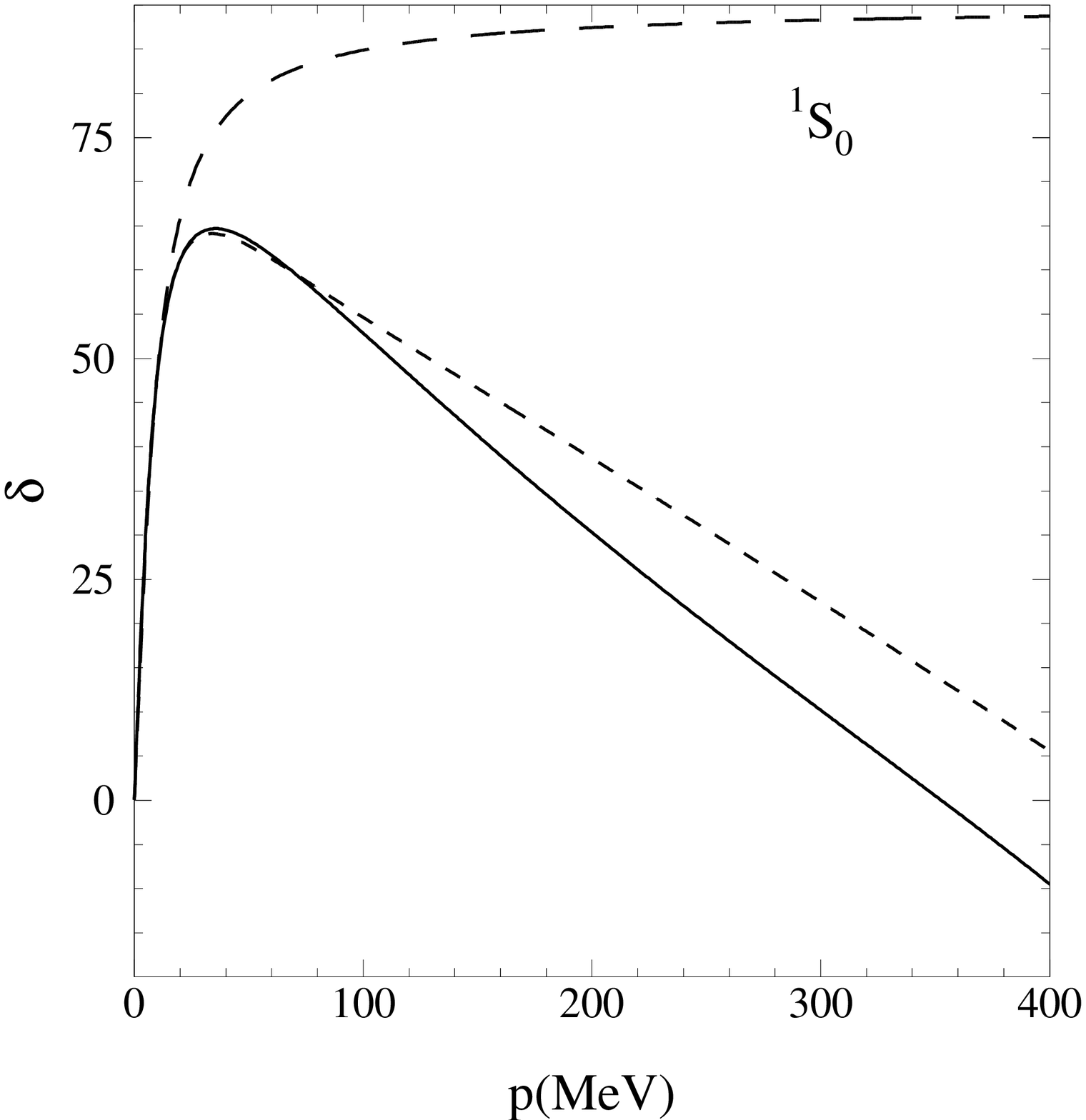}
        \epsfxsize=8truecm \epsfbox{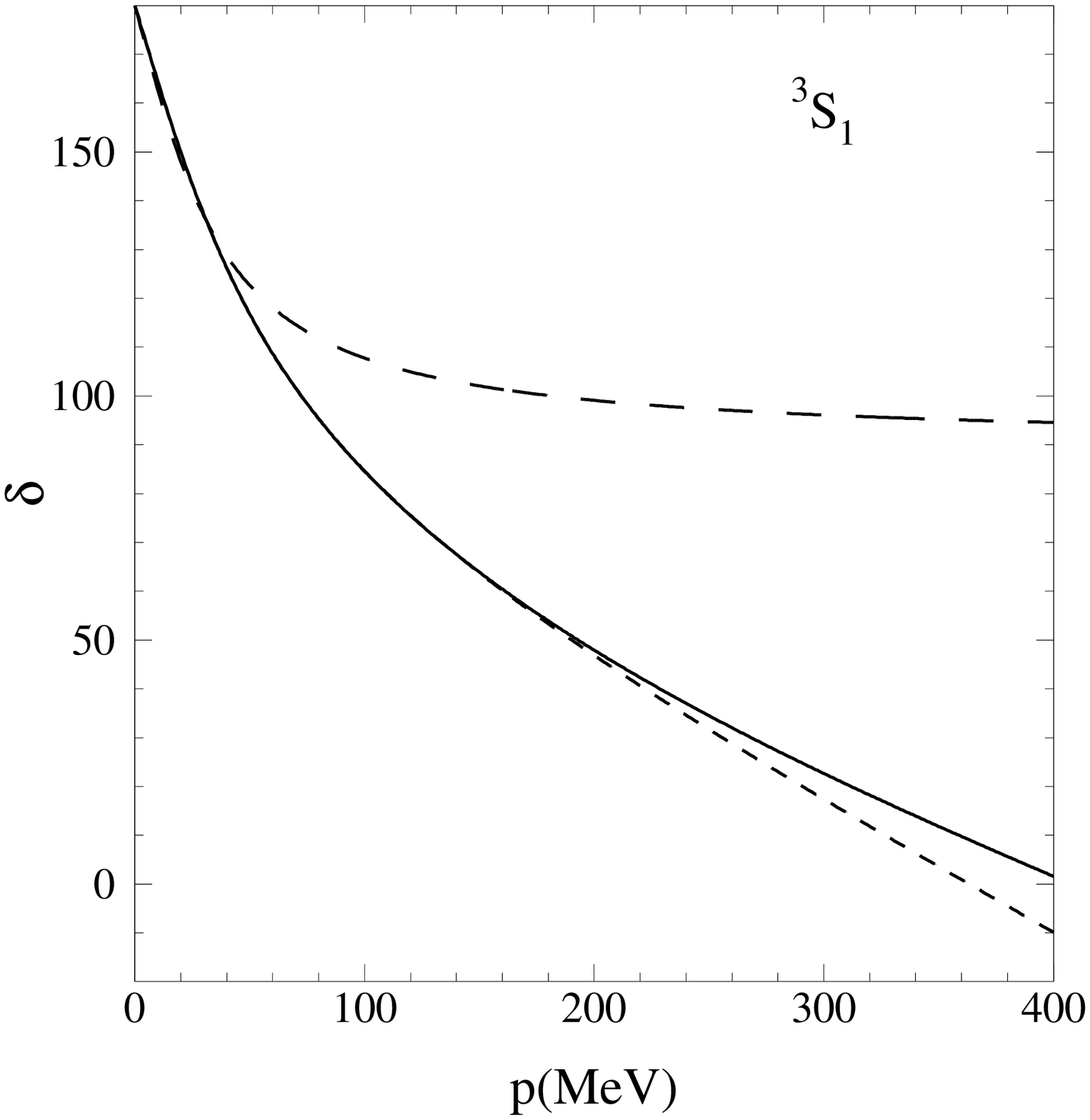} }
 {\tighten  
\caption[1]{Fit to the phase shift data emphasizing the low momentum 
region.  The solid line is the Nijmegen fit to the data \cite{Nij}, the long dashed 
line is the order $1/Q$ result, and the short dashed line is the order $Q^0$ 
result. } \label{fig_fits} }
\end{figure}

Using the Nijmegen phase shifts \cite{Nij} between $7$ and $100\,{\rm MeV}$, we fit
the coefficients $C_0(\mu_R)$, $C_2(\mu_R)$ and $D_2(\mu_R)$.  We took
$m_\pi=137\,{\rm MeV}$. Clearly we would like to bias the fit towards the low
momentum points since that is where the theoretical error is smallest.  This can be
accomplished by assigning a percent error, $\simeq p/(300\,{\rm MeV})$, to the data
and then minimizing the $\chi^2$ function.  In Tables~\ref{tble_sPDS} and
\ref{tble_sOS} we show the values\footnote{\tighten The coefficients extracted
from our fits differ from those in Ref.\protect\cite{ksw2} because we have
emphasized the low energy data as opposed to doing a global fit.  It is interesting
to note that using our PDS value $C_2(\mu_R =137\,{\rm MeV})=11.5\,{\rm fm^4}$,
the prediction for the RMS charge radius of the deuteron \protect\cite{ksw3}
becomes $1.966\,{\rm fm}$ which is within 1\% of the experimental result.} of
$C_0(\mu_R)$, $C_2(\mu_R)$ and $D_2(\mu_R)$ extracted from the fits for $\mu_R
= 70,100,137,160,280\,{\rm MeV}$.  These values exhibit the $\mu_R$ dependence
predicted by the RGE's to $\sim1\%$ in the $^1S_0$ channel and $\sim 4\%$ in the
$^3S_1$ channel.  In Fig.~\ref{fig_fits} the results of the fits are shown.  The results
of the fits shown in the figure are identical in both schemes.  Higher order
corrections will give contributions to $\delta$ of the form $p^2/\Lambda_\pi^2$. 
The error in $\delta$ at $p=300\,{\rm MeV}$ is consistent with $\Lambda_\pi \gtrsim
500\,{\rm MeV}$.

For processes involving the deuteron it is convenient to fix $C_0(\mu_R)$ using
the deuteron binding energy, $C_0(m_\pi)=-5.708\, {\rm fm}^2$.  With this
constraint we find $C_2(m_\pi)=10.80\,{\rm fm^4}$ and $D_2(m_\pi)=1.075\,{\rm
fm^4}$ in the PDS scheme. The fit to the phase shift data with these values is as
good as that in Fig.~\ref{fig_fits}.  

{\tighten \begin{table}[!t]
\begin{center} \begin{tabular}{ccccccccccc} 
& & \multicolumn{3}{c}{Fit to ${}^1\!S_0$} &&  
	\multicolumn{3}{c}{Fit to ${}^3\!S_1$} &\\ 
$\mu_R({\rm MeV}) $  && $C_0(\mu_R)$ & $C_2(\mu_R)$ & $D_2(\mu_R)$ & 
    & $C_0(\mu_R)$ & $C_2(\mu_R)$ &  $D_2(\mu_R)$ &  \\ \hline
$70$  &&  $-6.48$ & $10.11$ & $-0.532$    &&  
	 $-22.73$ & $171.$  & $-70.41$  &   \\ 
$100$  &&  $-4.71$ & $5.36$ & $1.763$    && 
	$-9.93$ & $32.7$ & $-4.157$ &  \\ 
$137$ &&  $-3.53$ & $3.01$ & $2.000$   && 
	$-5.88$ & $11.5$ & $1.500$ &  \\ 
$160$  && $-3.05$ & $2.25$ & $1.869$   && 
	$-4.69$ & $7.32$ & $1.897$ &  \\ 
$280$  &&  $-1.79$ & $0.772$ & $1.105$   && 
	$-2.19$ & $1.57$ & $1.004$ &  
\end{tabular} \end{center} 
{\tighten \caption{$^1S_0$ and $^3S_1$ couplings in the PDS
scheme.  $C_0(\mu_R)$ (in ${\rm fm}^2$), $C_2(\mu_R)$ (in ${\rm fm}^4$), and 
$D_2(\mu_R)$ (in ${\rm fm}^4$) are fit to the
Nijmegen data at different values of $\mu_R$.  } \label{tble_sPDS} } 
\end{table} } 

{\tighten
\begin{table}[!t]  
\begin{center} 
\begin{tabular}{ccccccccccc} & &
\multicolumn{3}{c}{Fit to ${}^1\!S_0$} &&  \multicolumn{3}{c}{Fit to ${}^3\!S_1$} &\\
  $\mu_R({\rm MeV}) $  && $C_0(\mu_R)$ & $C_2(\mu_R)$ & $D_2(\mu_R)$ &
  &  $C_0(\mu_R)$ & $C_2(\mu_R)$ &  $D_2(\mu_R)$ & \\   \hline 
$70$  && $-6.50$ & $9.75$ & $-6.047$   &&  
	 $-24.1$ & $121.$ & $-170.1$ &   \\ 
$100$  && $-4.73$ & $5.33$ & $-1.143$    && 
	$-10.0$ & $27.3$ & $-20.18$ &  \\ 
$137$ &&   $-3.54$ & $3.00$ & $0.378$  && 
	$-5.92$ & $10.5$ & $-4.124$ &  \\ 
$160$  && $-3.06$ & $2.25$ & $0.658$  && 
	$-4.74$ & $6.89$ & $-1.671$ &  \\ 
$280$  &&  $-1.80$ & $0.779$ & $0.692$  && 
	$-2.23$ & $1.61$ & $0.2985$ &  
\end{tabular}  
\end{center} 
{\tighten \caption{$^1S_0$ and $^3S_1$ couplings in the OS scheme. 
$C_0(\mu_R)$ (in ${\rm fm}^2$), $C_2(\mu_R)$ (in ${\rm fm}^4$), and 
$D_2(\mu_R)$ (in ${\rm fm}^4$) are fit to the Nijmegen data at
different values of $\mu_R$.  }
\label{tble_sOS} } 
\end{table} }

In PDS, it is necessary to break $C_0(\mu_R)$ into perturbative and
non-perturbative parts to obtain amplitudes that are $\mu_R$ independent
order-by-order.  If $C_0^p(\mu_R)$ is omitted then the values of $D_2(\mu_R)$
determined from the fit will not follow the RGE. To see this we define the $\mu_R$
independent quantity
\begin{eqnarray} 
  R &=& c \bigg[ \frac{-D_2(\mu_R)}
  {C_0(\mu_R)^2} + \frac{M}{8\pi} 
   \bigg(\frac{M g_A^2}{8\pi f^2}\bigg) \ln{\bigg({\mu_R^2 \over  \mu_0^2}
   \bigg)} \bigg] \,, 
\end{eqnarray} 
and choose the constant $c$ so that $R=1$ for $\mu_R=137\,{\rm MeV}$.  For
other values of $\mu_R$ the deviation of $R$ from $1$ gives the discrepancy
between the values predicted by the RGE and those extracted from the fit.  For
$\mu_R=70,280\,{\rm MeV}$ we find $R=-0.53,\,7.25$ in the $^1S_0$ channel and
$R=-0.52,\,11.4$ in the $^3S_1$ channel.  These large deviations disappear if
$C_0^p(\mu_R)$ is included.  Without $C_0^p(\mu_R)$, the PDS amplitude is still
$\mu_R$ independent to the order that one is working.  However, as explained
below, this residual $\mu_R$ dependence gives larger corrections than expected 
\cite{Gegelia2} since it makes the tuning that was setup to give the large 
scattering length $\mu_R$ dependent.

For momenta $p \ll m_\pi$ the pion can be integrated out leaving the effective
range expansion
\begin{eqnarray}
  p\cot{(\delta)} =  -\frac{1}{a} + \frac{r_0}{2}\, p^2 + 
     v_2\, p^4 + v_3\, p^6 + v_4\, p^8 + \ldots \,.
\end{eqnarray}
Performing a matching calculation between the two theories gives expressions
for $a$, $r_0$ and the $v_i$ in terms of the parameters in the pion theory. 
Since the theory with pions is an expansion in $Q$ these predictions take the 
form of Taylor series in $Q/\Lambda_\pi$
\begin{eqnarray}  \label{erce}
  \frac1a = \gamma +  \sum_{i=2}^\infty B_0^{(i)} \,, \qquad\quad 
   \frac{r_0}{2} = \sum_{i=0}^\infty  B_1^{(i)} \,, \qquad\quad
    v_n = \sum_{i=2-2n}^\infty  B_n^{(i)} \,.
\end{eqnarray}
where $B_n^{(i)} \sim Q^{\,i}$.  At this time only the first coefficient in each series
is known since $p\cot\delta$ has only been calculated to order $Q^2$.  The 
notation
\begin{eqnarray}  \label{C0g}
 \gamma = \frac{4\pi}{M C_0(\mu_R)} + \mu_R \,
\end{eqnarray}
will be used to denote the location of the perturbative pole in the amplitudes. In 
PDS
\begin{eqnarray} \label{B01PDS}
  B_0^{(2)} &=& \bigg( \frac{-4\pi }{M} \bigg) {m_\pi^2 D_2(\mu_R)   \over 
    C_0(\mu_R)^2} +\frac{m_\pi^2\,M g_A^2}{8\pi f^2}
    \Bigg[ \frac12 + \frac12 \ln{\bigg({\mu_R^2 \over
    m_\pi^2}\bigg)} -\frac{2\gamma}{m_\pi}  
  +\frac{\gamma^2}{m_\pi^2} \ \Bigg]   - K \nn \,, \\
 B_1^{(0)} &=& \bigg( \frac{4\pi}{M} \bigg) 
  \frac{C_2(\mu_R)}{\Big[C_0(\mu_R) \Big]^2} + \frac{Mg_A^2}{8\pi f^2} 
  \bigg[  1 - \frac{8\gamma}{3\,m_\pi} +   \frac{2\gamma^2}{ m_\pi^2 } \bigg]  \,.
\end{eqnarray}
Note that if $C_0^p(\mu_R)$ had been neglected then $B_0^{(2)}$ would not be
$\mu_R$ independent.  With  $\mu_R=m_\pi$ Eq.~(\ref{B01PDS}) agrees with 
Ref.~\cite{ksw2} if their definition of $D_2(\mu_R)$ is adopted. In the OS scheme 
we have 
\begin{eqnarray}  \label{B01OS}
  B_0^{(2)} &=& \bigg( \frac{-4\pi}{M}\bigg)  { m_\pi^2 D_2(\mu_R)\over 
  C_0(\mu_R)^2} +\frac{ m_\pi^2 \,Mg_A^2}{8\pi f^2}
    \bigg[ \frac12 \ln{\bigg({\mu_R^2 \over
    m_\pi^2}\bigg)} -\frac{2\gamma}{m_\pi}  \bigg]  -K \,, \nn \\
 B_1^{(0)} &=&  \bigg( \frac{4\pi}{M} \bigg) 
  \frac{C_2(\mu_R)}{C_0(\mu_R)^2} + \frac{Mg_A^2}{8\pi f^2} 
  \bigg[ \frac{\gamma^2}{\mu_R^2 } +1- \frac{8\gamma}{3\,m_\pi} + \frac{2\gamma^2}
  {m_\pi^2} \bigg]  \,.
\end{eqnarray}
The value of the remaining $B_n^{(i)}$ determined at this order are the same in
both schemes 
\begin{eqnarray}  \label{Bvn}
  B_n^{(2-2n)} &=& - \frac{M g_A^2}{8\pi f^2} \left(\frac{-4}{m_\pi^2}\right)^n 
  \bigg[\frac{1}{4n} - \frac{2\gamma}{(2n+1)m_\pi} + \frac{\gamma^2}{(n+1)m_\pi^2}
    \bigg]\, m_\pi^2 \,.
\end{eqnarray}
For $n=2,3,4,$ Eq.~(\ref{Bvn}) gives the low-energy theorems derived in
Ref.~\cite{Cohen2} if we set $\gamma=1/a$.

{\tighten
\begin{table}[!t] 
\begin{center} 
\begin{tabular}{ccccccccccc} & &
\multicolumn{4}{c}{${}^1\!S_0$ Fit}  && \multicolumn{4}{c}{${}^3\!S_1$ Fit}  
  \vspace{.05in} \\ 
  $\mu_R({\rm MeV}) $  && $\gamma$ & $B_0^{(2)}$ & $1/a$ &
  $r_0$ &&  $\gamma$ & $B_0^{(2)}$ & $1/a$ &  $r_0$  \\   \hline 
 $70$  && $-10.18$ & $2.05$ & $-8.124$ & $0.01468$ && 
	$48.39$ & $-15.82$ & $32.57$ & $0.01101$ \\ 
$137$ && $-10.16$ & $2.04$ & $-8.121$ & $0.01480$ &&
	$48.96$ & $-16.76$ & $32.19$ & $0.01098$ \\ 
$280$  && $-10.23$ & $2.12$ & $-8.105$ & $0.01484$ &&
	$46.39$ & $-12.64$ & $33.76$ & $0.01111$
\end{tabular}  
\end{center} 
{\tighten \caption{Values of $\gamma$, $B_0^{(2)}$, $1/a$, and $r_0$ (in MeV) 
obtained from our fits. Three values of $\mu_R$ are shown to emphasize that the 
value of the extracted parameters depends weakly on $\mu_R$.}
\label{tble_ar0} } 
\end{table} }
Recall that the unnaturally large scattering length $a$ is a fine tuning that was
accounted for by demanding that in Eq.~(\ref{C0g}), $C_0(\mu_R)$ is close to its
ultraviolet fixed point, and $\gamma\approx 1/a$.  Examining the expression for
$1/a$ in Eq.~(\ref{erce}) it may seem that this could be destroyed by chiral
corrections.  If $D_2(\mu_R)\sim C_0(\mu_R)^2$ then the first term gives
$B_0^{(2)}\sim 205\,{\rm MeV}$.  In fact from Table~\ref{tble_ar0}, we see that the fit
gives $B_0^{(2)}\lesssim 1/a$.  The reason for this small value is that since
$A^{(0)} \propto (A^{(-1)})^2$ the amplitude has a double pole.  Since this pole is
spurious (occurring from the perturbative expansion) the residue of the double
pole must be small in order to fit the data.  This leads to a good fit
condition\cite{ms0} which will be approximately satisfied 
\begin{eqnarray}
     \left.  \frac{A^{(0)}}{[A^{(-1)}]^2} \ \right|_{-ip = \gamma} = 0\,.
\end{eqnarray}
As explained in Ref.~\cite{ms0}, this condition implies $B_0^{(2)}\simeq 
4\pi\gamma^2/M$.  In fact this gives the right order of magnitude 
for the values of $B_0^{(2)}$ determined from the fits in Table~\ref{tble_ar0}.
Similar good fit conditions occur at higher order keeping the coefficients
$B_0^{(i)}$ small.  Thus the tuning $\gamma\approx 1/a$ is not destroyed, but
instead naturally kept by the form of the perturbative expansion.  The division
of $C_0(\mu_R)$ into nonperturbative and perturbative pieces is arbitrary, allowing
us to set up the theory so that the $Q$ expansion for $1/a$ is well behaved.

In Table~\ref{tble_ar0} we see that when $B_0^{(2)}$ is added to $\gamma$, values
of $1/a$ are obtained which are close to the physical values, $1/a(^1S_0)=
-8.32\,{\rm MeV}$ and $1/a(^3S_1)=36.4\,{\rm MeV}$.  It is encouraging that the
value of $\gamma$ found from fits in the $^3S_1$ channel are close to the physical
pole in the amplitude which corresponds to the deuteron, $\gamma=45.7\,{\rm
MeV}$.  Values for $r_0$ can also be predicted from the fits using
Eq.~(\ref{B01OS}).  Experimentally, $r_0(^1S_0)=0.0139\,{\rm MeV}^{-1}$ and
$r_0(^3S_1)=0.00888\,{\rm MeV}^{-1}$, so the values in Table~\ref{tble_ar0} agree
to the expected accuracy.  It is not yet clear\cite{ms0} whether values of the $v_i$
extracted from experimental data \cite{Cohen2,Nij2} are accurate enough to test
the low-energy theorems for $v_2$, $v_3$, and $v_4$.

\section{Determining the range $\Lambda_\pi$ }

Here we will examine the phase shift data to see what it tells us about the range of
the effective field theory with perturbative pions.  In Ref.~\cite{sf2}, a Lepage
analysis is performed on the observable $p\cot{\delta(p)}$ in the $^1S_0$ channel. 
Near $350\,{\rm MeV}$ the experimental $^1S_0$ phase shift passes through zero. 
Therefore, the error $|p\cot{\delta}^{\rm NPWA}-p\cot{\delta}^{\rm EFT}|$ is greatly
exaggerated since $p\cot\delta(p)\to \infty$. To avoid this problem we will use the
$^1S_0$ and $^3S_1$ phase shifts as our observables, since $\Delta \delta=
|\delta^{\rm NPWA}-\delta^{\rm EFT}|$ remains finite for all $p$.  The
next-to-leading order amplitudes given in section V will be used. The phase shifts
have an expansion of the form $\delta = \delta^{(0)} + \delta^{(1)} + {\cal
O}(Q^2/^2)$, where \cite{ksw2} 
\begin{eqnarray}
  \delta^{(0)} &=& -\frac{i}{2} \ln{\bigg[ 1 + i\frac{pM}{2\pi} {\cal A}^{(-1)}
	\bigg]} \,,\qquad
   \delta^{(1)} = \frac{pM}{4\pi} \frac{ {\cal A}^{(0)} }{ 1 + i\frac{pM}{2\pi}
  {\cal A}^{(-1)} }\,.  \label{delta01}
\end{eqnarray}
Recall that a momentum expansion of $\delta$ would result in terms with only odd
powers of $p$.  However, the expansion for $\delta$ in Eq.~(\ref{delta01}) is not
simply a momentum expansion, so the next-to-leading order calculation can have
errors which scale as $p^2/\Lambda_\pi^2$.  For example, once pions are included we
can have a term $p^2 \tan^{-1}(2 p/m_\pi)$ which is odd in $p$, order $Q^2$, and
scales as $p^2$ for large momenta. 

\begin{figure}[!t]  
  \centerline{\epsfxsize=8.0truecm \epsfbox{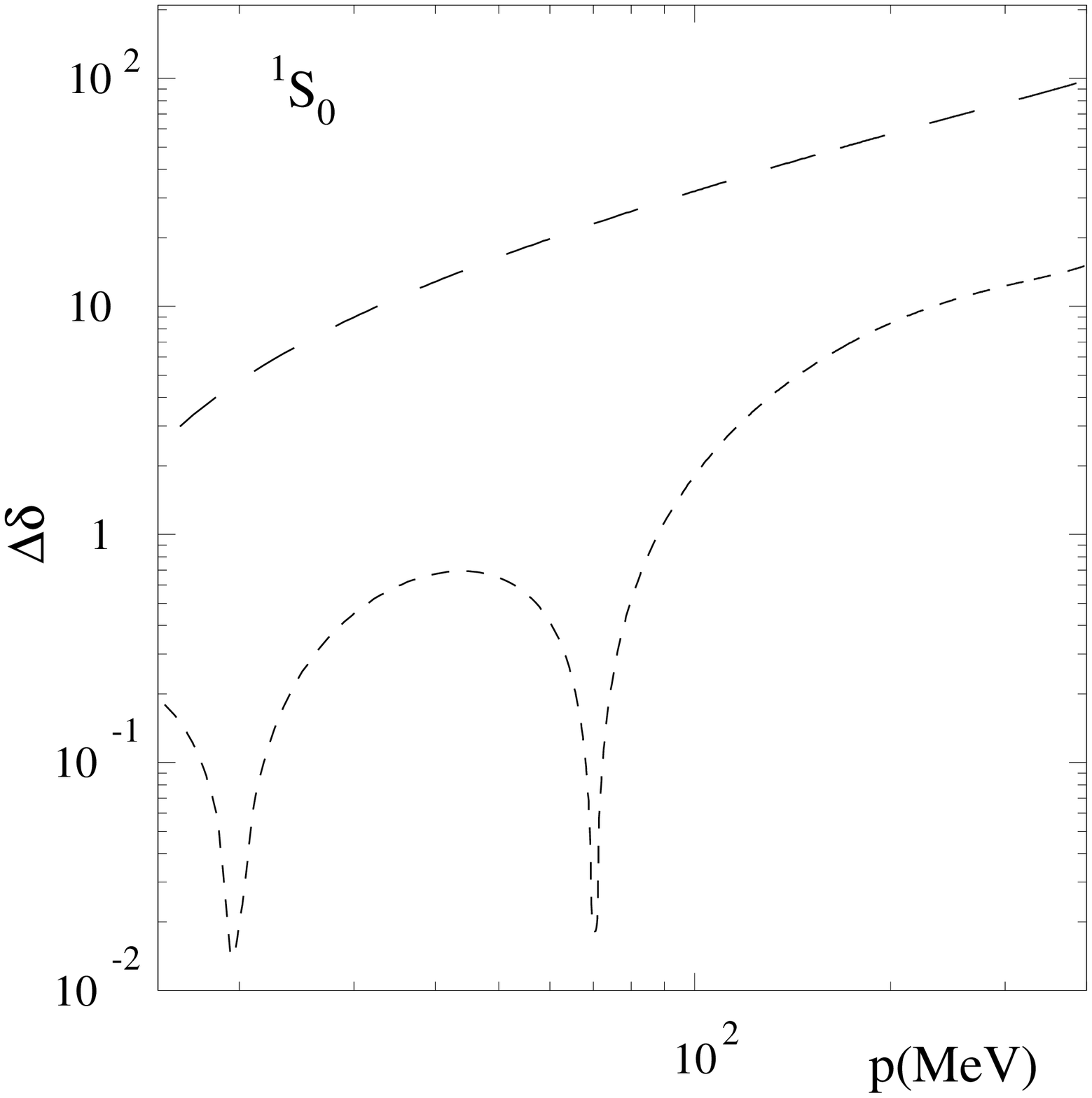}
        \epsfxsize=8truecm \epsfbox{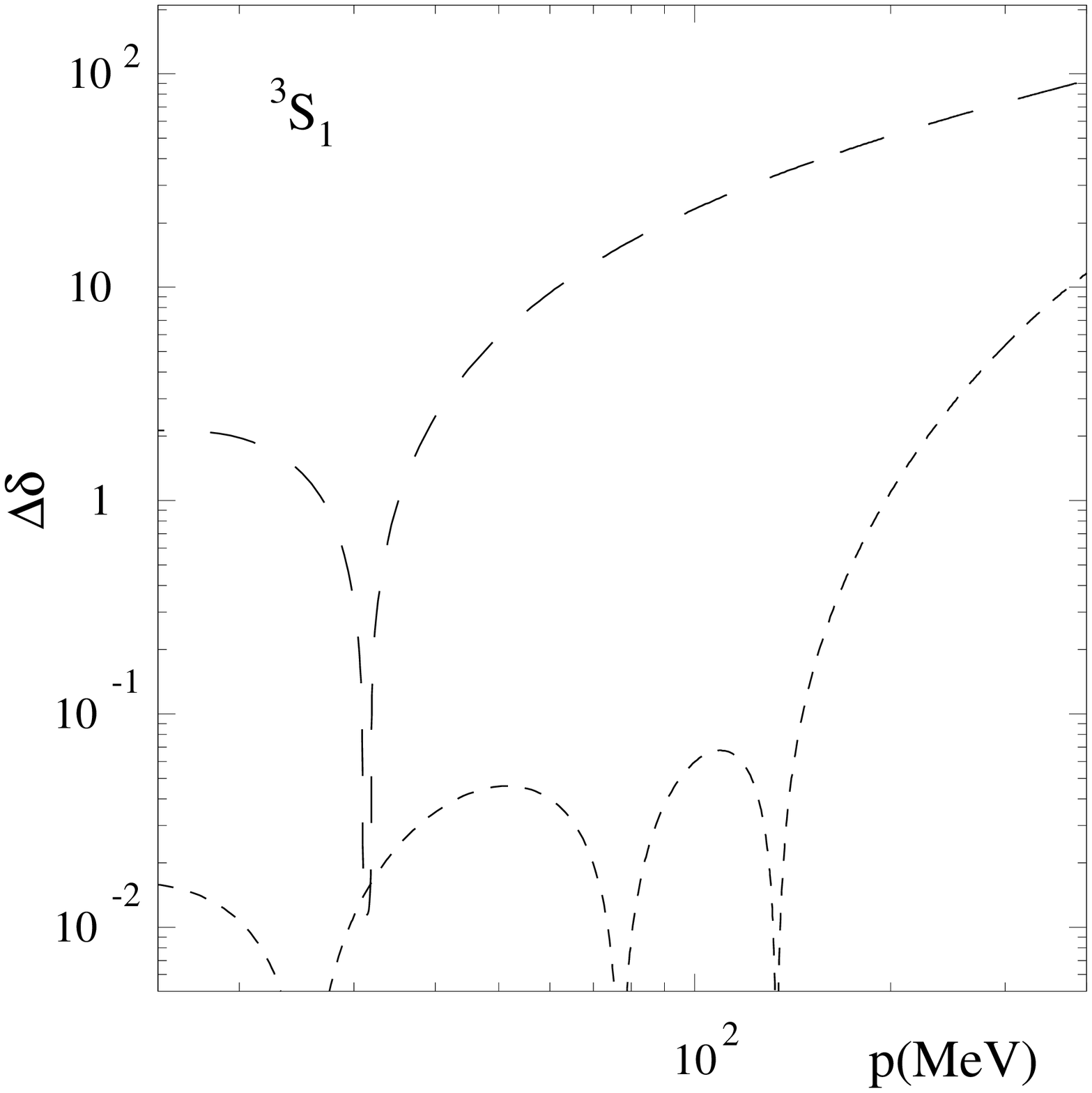} }
 {\tighten  
\caption[1]{Error analysis for the phase shifts in the ${}^1\!S_0$ and
${}^3\!S_1$ channels.  $\Delta \delta$ is the difference between the
the effective field theory prediction and the Nijmegen partial wave 
analysis\cite{Nij}. 
The long and short dashed lines use the ${\cal O}(Q^0)$ and ${\cal O}(Q)$ 
theoretical phase shifts respectively.} \label{fig_lep} }
\end{figure}

In Fig.~\ref{fig_lep}, we plot $\Delta \delta$ versus $p$ using log-log axes.  Note
that the sharp dips in Fig.~\ref{fig_lep} are just locations where the theory happens
to agree with the data exactly.  The Nijmegen data\cite{Nij} is available up to
$p=405\,{\rm MeV}$.  
In a theory with just a momentum expansion the errors will appear as straight lines
on the log-log plot as pointed out by Lepage \cite{Lepage}.  In the pion theory the
expansion is in both $m_\pi/\Lambda_\pi$ and $p/\Lambda_\pi$, so this is no longer
true.   For $p>m_\pi$ we expect the errors to be of the form\footnote{\tighten At
momenta $1/a \ll p \ll m_\pi$ we could have $\Delta \delta^{(0)} \sim
B_0^{(2)}/p\sim m_\pi^2/\Lambda_\pi p$.  However, as explained in section VI,
$B_0^{(2)}\lesssim 1/a$ so this term is very small.}
\begin{eqnarray}
 \Delta \delta^{(0)} &\sim& \bigg(1+\frac{m_\pi}{\Lambda_\pi}+\ldots\bigg) {p\over
    \Lambda_\pi} +\ldots\,, \\ \Delta \delta^{(1)} &\sim&
   \bigg(\frac{m_\pi}{\Lambda_\pi}+\frac{m_\pi^2}{\Lambda_\pi^2}+ \ldots \bigg)
  {p\over \Lambda_\pi} + \bigg(1+\frac{m_\pi}{\Lambda_\pi}+\ldots \bigg) 
  {p^2\over \Lambda_\pi^2} +\ldots \nn \,.
\end{eqnarray}
The fact that there is always a $p/\Lambda_\pi$ error arises from the fact that, as
seen in Eq.~(\ref{erce}), $r_0$ is reproduced in the effective field theory as an
expansion in $m_\pi/\Lambda_\pi$.  For $p/\Lambda_\pi \gg m_\pi/\Lambda_\pi$ the
slope of the lines on the plot should indicate the lowest power of $p$ that has not
been included.  At low momentum the error in $\Delta \delta^{(n)}$ is dominated by
the $p\, m_\pi^n/\Lambda_\pi^{n+1}$ term and the lines should be parallel.  From
Fig.~\ref{fig_lep} we see that the error is smallest at low momentum and increases
as the momentum increases, which is how the theoretical error is expected to
behave.  

It is clear that even for $p \sim 400\,{\rm MeV}$ the next-to-leading order
calculations are reducing the error in the phase shift.  Because two new
parameters are added at next-to-leading order it is always possible to force exact
agreement at some value of p.  However, if one were to force the data to agree too
well at high momentum then this would destroy the agreement at low momentum. 
Since the improvement of the fit in Fig.~\ref{fig_lep} at high momentum does not
come at the expense of the fit at low momentum this is evidence that the error is
being reduced in a systematic way.  At high momentum one expects that the error
is $\sim p^2/\Lambda_\pi^2$.  From Fig.~\ref{fig_lep}, $\Delta \delta\sim 0.26\,{\,\rm
radians}$ for $p=400\,{\rm MeV}$, implying $\Lambda_\pi\sim 800\,{\rm MeV}$.  This
is only a rough estimate for the range because we cannot yet exclude the
possibility that the next-to-next-to-leading order phase shift has an anomalously
small coefficient.  Even though the lines in Fig.~\ref{fig_lep} are not straight they
should still cross at approximately the range of the theory since at this point
higher order corrections do not improve the agreement with the data.   This error
analysis is consistent with the possibility that the range is $\gtrsim 500\,{\rm
MeV}$.  

Further information on the range of the effective field theory can be obtained by
examining electromagnetic processes involving the deuteron \cite{ksw3,metal},
such as the deuteron charge radius, electromagnetic form factors, deuteron
polarizability, and deuteron Compton scattering.  For these observables errors are
typically $\sim 30-40\%$ at leading order and $\sim 10\%$ at next-to-leading
order.  This is what one would expect if the expansion parameter
$m_\pi/\Lambda_\pi\sim 1/3$, implying $\Lambda_\pi\sim 410\,{\rm MeV}$.  This is
consistent with our previous estimate for $\Lambda_\pi$.  If the range is this large
one should expect that the error in deuteron properties will be at the few percent
level once  next-to-next-to-leading order calculations are performed.  

\section{Conclusion}

In this paper the structure of the effective field theories of nucleons with and
without pions is studied.  We discuss a momentum subtraction scheme, the OS
scheme, which obeys the KSW power counting.  The method of local counterterms
is used to obtain the renormalization group equations for the coupling constants in
these theories.  Using local counterterms defines the OS and PDS renormalization
schemes unambiguously.  Two-loop graphs with potential pions in the $^3S_1$
channel are computed and shown to have $p^2/\epsilon$ poles.  The presence of
$1/\epsilon$ poles implies that the only model independent piece of pion exchange
is the part that can be treated perturbatively.  We obtain the renormalized
couplings $C_0(\mu_R)$, $C_2(\mu_R)$ and $C_4(\mu_R)$ at order Q in the OS
and PDS schemes.  

We have emphasized why it is important to have $\mu_R$ independent amplitudes
order by order in $Q$.  Such amplitudes are obtained automatically in the OS
scheme.  In PDS $\mu_R$ independent amplitudes may be obtained by treating
part of $C_0(\mu_R)$ perturbatively.  Another result concerns the large $\mu_R$
behavior of the couplings in this theory. In the OS scheme the coupling constants
obey the KSW power counting for all $\mu_R>1/a$.  In PDS the breakdown in the
power counting for $C_0(\mu_R)$ is avoided if $C_0(\mu_R)$ is split into
non-perturbative and perturbative parts.  Therefore, the breakdown of the scaling
in the coupling constants is artificial.

Next-to-leading order calculations of nucleon-nucleon phase shift data
\cite{ksw2} provide fits to data at large momenta which are far more accurate than
one would expect if the theory broke down completely at $300 {\rm\,MeV}$.  Of
course, this does not mean that nucleon effective theory can be applied at
arbitrarily high energies.  The scale, $M g_A^2/(8 \pi f^2) \sim 300\,{\rm MeV}$, is
associated with short distance contributions from pion exchange and provides an
order of magnitude estimate for the range.  In the S-wave channel, $\Delta$
production and $\rho$ exchange become relevant at $\sim 700\,{\rm MeV}$, which
sets an upper limit on the range of the expansion.  To get a better understanding of
the range of the nucleon effective theory with perturbative pions one must examine
experimental data. An error analysis of the S wave phase shifts with
next-to-leading-order calculations seems to be consistent with a range of
$500\,{\rm MeV}$.  Though next-to-next-leading order corrections need to be
compared with data and other processes investigated, we remain cautiously
optimistic that the range could be as large as $500\,{\rm MeV}$.

\acknowledgments

We would like to thank Mark Wise for many useful conversations.  We also would
like to thank H. Davoudiasl, S. Fleming, U. van Kolck, Z. Ligeti, S. Ouellette and
K.  Scaldeferri, for their comments.  T.M. would like to acknowledge the
hospitality of the Department of Physics at the University of Toronto, where part
of this work was completed.  This work was supported in part by the Department
of Energy under grant number DE-FG03-92-ER 40701.  T.M. was also supported
by a John A. McCone Fellowship.


\appendix
\section{Loop integrals with a momentum cutoff regulator}

Although the analysis in section III used dimensional regularization to regulate
divergent loop integrals, the results for the coefficients $C_{2m}(\mu_R)$ in
our momentum subtraction scheme are independent of this choice.  As an
exercise we will derive the counterterms for $C_0(\mu_R)$ and $C_2(\mu_R)$
using a momentum cutoff regulator, $\Lambda$.  This will give us the chance
to see what type of complications can arise using a different regulator.  Note
that this is not the same as using a {\em{finite}} cutoff scheme.  There the
momentum cutoff plays a double role as both a regulator and as part of the
subtraction scheme.

The graph in the first row first column of Fig.~\ref{fig_ct0} gives
\begin{eqnarray}
  i C_0^2 & & M  \int_0^\Lambda \frac{d^3 q}{(2\pi)^3} 
     \frac1{\vec q\,^2-p^2}     \nn \\ 
&&  = \frac{iM}{2\pi^2} C_0^2 \bigg[ \Lambda + \frac{i\pi p}2 
     - p {\rm tanh}^{-1}{\left(\frac{p}{\Lambda}\right)} \bigg]   \\
&&  = \frac{iM}{2\pi^2} C_0^2 \bigg[ \Lambda + \frac{i\pi p}2
     - \frac{p^2}{\Lambda} -\frac{p^4}{3\Lambda^2} -\ldots \bigg]   \nn  \,.
\end{eqnarray}
An ultraviolet counterterm cancels the linear divergence,
\begin{eqnarray}
 \delta^{1,{\rm uv}}C_0 = \frac{M}{4\pi} {C_0^{\rm finite}}^2 \left( - 
   \frac{2\Lambda}{\pi}  \right) \,. \label{dc0L}
\end{eqnarray}
and the same finite counterterm, $\delta^1C_0(\mu_R)$ in 
Eq.~(\ref{ct024}) 
is used to satisfy the condition in Fig.~\ref{fig_C0}.  The renormalized graph 
is then the same as calculated in dimensional regularization in section III.
Note that contributions of order $p^2$ have been neglected in defining 
$C_0(\mu_R)$ as required by our renormalization condition.  An
added complication with a cutoff is that graphs with only $C_0$'s give a 
contribution to the amplitude proportional to $p^2$.  However, as 
$\Lambda \to \infty$, $p\:{\rm tanh}^{-1}{(p/\Lambda)}\to 0$, so these terms
can be completely neglected.  This will remain true even for higher loops 
since the counterterms will always cancel dangerous powers of $\Lambda$
that appear in the numerator.  At $n$ loops we find an ultraviolet 
counterterm of the form
\begin{eqnarray}
  \delta^{n,{\rm uv}}C_0 = -\bigg(\frac{-M}{4\pi}\bigg)^n C_0(\mu_R)^{n+1} 
    \left(  - \frac{2\Lambda}{\pi} \right)^n \,,  \label{dc0nL}
\end{eqnarray}
while the finite counterterms are given by Eq.~(\ref{ct024}).

The graph in the third row first column of Fig.~\ref{fig_ct0} gives
\begin{eqnarray}
  2 i C_0 & & C_2 \frac{M}2 \int_0^\Lambda \frac{d^3 q}
    {(2\pi)^3} \frac{\vec q\,^2+p^2}{\vec q^2-p^2}   \nn \\ 
&&  = 2\,\frac{iM}{2\pi^2} C_0 C_2 \left\{ \frac{\Lambda^3}{6} 
  + p^2\left[ \Lambda +  \frac{i\pi p}2 - p {\rm tanh}^{-1}
  {\left(\frac{p}{\Lambda}\right)} \right]  \right\} \,. \label{c0c2L}
\end{eqnarray}
Note that there are different contributions from this graph when the
vertices are in the order $C_0\,C_0\,C_2$ or $C_0\,C_2\,C_0$.
At order $p^2$, this graph gives a correction to the counterterm
$\delta^{1,{\rm uv}}C_0$, i.e., $\delta^{1,{\rm uv}}C_0 \to \delta^{1,{\rm uv}}
C_0 + \delta^{1*,{\rm uv}}C_0$, where
\begin{eqnarray}
\delta^{1*,{\rm uv}}C_0 &=& - \frac{M}{4\pi} 2\, C_0^{\rm finite}\, C_2^{\rm finite}
    \frac{\Lambda^3}{3\pi} \,. 
\end{eqnarray}
Unlike the contribution to $\delta^{1,{\rm uv}}C_0$ in Eq.~(\ref{dc0nL}), 
$\delta^{1*,{\rm uv}}C_0$ is to be treated perturbatively, so that at it only 
appears once in any graph.  The justification of this fact is 
that this contribution to the counterterm appeared at order $Q^0$ (a purely 
formal trick to recover this counting is to take $\Lambda\sim \mu_R \sim Q$).  
The counterterm $\delta^{1,{\rm uv}}C_2$ is fixed by considering the order 
$p^2$ terms in Fig.~\ref{fig_ct0}, row 3.  From Eq.~(\ref{c0c2L}) 
(the ${\rm tanh}^{-1}$ piece can again be thrown away) we have
\begin{eqnarray}
 \delta^{1,{\rm uv}}C_2 &=& \frac{M}{4\pi} 2\, C_0^{\rm finite}\, C_2^{\rm finite} 
   \left( - \frac{2\Lambda}{\pi}  \right)   \,.  \label{dC2L}
\end{eqnarray}
The calculation for higher loops is similar and there are again corrections 
$\delta^{n*,{\rm uv}} C_0$ to $\delta^{n,{\rm uv}} C_0$
\begin{eqnarray}
 \delta^{n*,{\rm uv}} C_0 &=&  \bigg(\frac{-M}{4\pi}\bigg)^n n(n+1)\, 
  ({C_0^{\rm finite}})^n\, C_2^{\rm finite} \left( - \frac{2\Lambda}{\pi} 
   \right)^{n-1} \frac{\Lambda^3}{3\pi} \,, \nn \\
 \delta^{n,{\rm uv}}C_2 &=& -\bigg(\frac{-M}{4\pi}\bigg)^n (n+1)\, 
  ({C_0^{\rm finite}})^n\, C_2^{\rm finite} \left( - \frac{2\Lambda}{\pi}  
  \right)^n   \,.  
\end{eqnarray}
The finite counterterms are the same as in Eq.~(\ref{ct024}).
Thus the running couplings and amplitudes with a cutoff are the same as
found using dimensional regularization.  

\section{Relations between couplings in OS and PDS}

Here we give explicit relations between the coupling constants that occur
at order $Q^0$ in the OS and PDS schemes.  Couplings on the left are in PDS 
while those on the right are in the OS scheme.	
\begin{eqnarray}
   C_0(\mu_R) &=& C_0(\mu_R) \,, \nn\\
   \frac{C_2(\mu_R)}{C_0(\mu_R)^2} &=& \frac{C_2(\mu_R)}{C_0(\mu_R)^2}  +
	\frac{g_A^2}{2f^2} \frac1{\mu_R^2} \bigg[ \frac1{C_0(\mu_R)} + 	
	\frac{M\mu_R}{4\pi} \bigg]^2 \,, \nn \\
  \frac{D_2(\mu_R)}{C_0(\mu_R)^2}-\frac{M}{8\pi} \bigg( \frac{Mg_A^2}{8\pi f^2} 
      \bigg)  &=& \frac{D_2(\mu_R)}{C_0(\mu_R)^2}  \,,  \\
   \frac{C_0^p(\mu_R)}{C_0(\mu_R)^2} + \frac{g_A^2}{2f^2} \frac1{C_0(\mu_R)^2}
	&=& \frac{C_0^p(\mu_R)}{C_0(\mu_R)^2} + \frac{g_A^2}{2f^2} 
	\bigg[ \frac1{C_0(\mu_R)} + \frac{M\mu_R}{4\pi} \bigg]^2 \nn \,.
\end{eqnarray}
As in section VI the superscript $np$ on $C_0(\mu_R)$ has been dropped.	

{\tighten

} 

\end{document}